\documentclass{aa}  

\usepackage{graphicx}
\usepackage{txfonts}
\usepackage{hyperref}
\begin{document}
  
\title{Phosphorus in cool stars of various metallicities: The non-local thermodynamic equilibrium consideration}

\titlerunning{Phosphorus in cool stars: non-LTE consideration}
\authorrunning{S.M. Andrievsky, S.A. Korotin}

\author{
S.M. Andrievsky\inst{1}
\and  
S.A. Korotin\inst{2}
}

\institute{
Astronomical Observatory, Odessa National University, Shevchenko Park, 65014 Odessa, Ukraine\\
\email{andrievskii@ukr.net}
        \and
Crimean Astrophysical Observatory, Nauchny 298409, Republic of Crimea\\
}

\date{Received ; accepted }

\abstract{}
{The phosphorus abundance distribution in field stars as a function of metallicity reveals a complex pattern. The local thermodynamic 
equilibrium (LTE) data for [P/Fe] in the low-metallicity range are sparse and scattered around [P/Fe] $\approx 0$ dex. Near [Fe/H] $\approx 
-2$ dex, the relative abundance [P/Fe] increases and reaches a maximum value of around [Fe/H] $\approx -1$ dex. In this domain, many 
phosphorus-rich (P-rich) stars and (super)phosphorus-rich stars are observed; the [P/Fe] value can exceed 1 dex. Until now, no attempts have 
been made to study the non-local thermodynamic equilibrium (non-LTE) effects on the ultraviolet and infrared phosphorus lines in spectra of 
cool stars to test the robustness of the observed LTE phosphorus abundance distribution.}  
{We developed an atomic model of \ion{P}{i} that can be used to analyze phosphorus lines in the spectra of cool dwarfs and giants in the 
non-LTE approximation. The model consists of 101 energy levels of \ion{P}{i} and the ground level 
of \ion{P}{ii}; 1070 transitions between mentioned levels were studied. The model was tested using the solar flux and intensity spectra, 
as well as the spectra of Procyon and $\sigma$~Boo. Profiles of 14 phosphorus lines in the infrared regions and equivalent widths were 
analyzed. Our non-LTE phosphorus abundance in the Sun is (P/H) = 5.35 $\pm 0.04$ dex.}
{Using our non-LTE model, we selected 12 ultraviolet and infrared phosphorus lines and calculated a grid of non-LTE corrections 
for the following parameter ranges: T$_{\rm eff}$ from  4000 to 6750 \,K, step 250 \,K; $\log~g$ from 1 to 5 dex, step 1 dex; and V$_{\rm t} = 
2$ km~s$^{-1}$, [Fe/H] from --3 to +0.5 dex, step 0.5 dex. The non-LTE corrections (Abundance$_{\rm non-LTE}$ -- Abundance$_{\rm LTE}$) 
were calculated for phosphorus abundance ratios of [P/Fe] = --0.4, 0.0, +0.4 dex. For the Sun, the non-LTE correction is --0.08 dex.} 
{The grid of the non-LTE corrections, as well as the direct line profile synthesis, were used to refine the literature data on the phosphorus
abundance in metal-poor, intermediate-deficient, and solar-metallicity stars. This sample also includes phosphorus-rich stars.
Non-LTE corrections do not qualitatively alter the overall phosphorus abundance distribution over a wide metallicity range, and 
do not change the characteristic pattern of phosphorus-rich stars. After corrections, the phosphorus abundance distribution became more 
compact in the low-metallicity range. Overall, the observed phosphorus abundance distribution can be described by the combined effect of 
phosphorus production in rotating massive stars, Type II supernovae explosions, and oxygen-neon-magnesium novae.}


\maketitle

\section{Introduction}

The importance of phosphorus in Cosmos is impossible to overestimate, given the biological structures existing on Earth, where 
phosphorus, along with hydrogen, carbon, nitrogen, oxygen, and sulfur, is one of the most essential components of the life. This 
chemical element is a part 
of the phosphate backbone of nucleotides in DNA, RNA (the phosphodiester backbone of their helical structure), the rechargeable energy 
“batteries” of the cells: ATP, ADF, AMP, and the phospholipides of cell membranes. Regardless of the origin of life and many purely scientific 
and philosophical questions concerning our understanding of how to resolve them, phosphorus nuclei are present in organic matter. How this 
element participated in the formation of life remains to be understood, but the fact that phosphorus nuclei were formed at certain points 
in the Universe is clear. In particular, this occurred in massive stars. 

The only stable isotope of phosphorus is ${}^{31}_{15}$P. In massive stars the phosphorus nuclei can form during the oxygen-burning 
phase in reactions between oxygen nuclei: 

${}^{16}_{8}$O(${}^{16}_{8}$O,${}^{1}_{1}$H)${}^{31}_{15}$P 

or 

${}^{16}_{8}$O(${}^{16}_{8}$O,$n$)${}^{31}_{16}$S($e^{+} \nu$)${}^{31}_{15}$P.

Another possibility for the formation of phosphorus nuclei is the proton capture by nuclei of the silicon isotope ${}^{30}_{14}$Si during 
the hydrostatic neon burning. These reactions occur in the progenitors of the Type II core-collapse supernovae. Theoretically,
phosphorus can be produced in asymptotic giant branch (AGB) stars. The $\beta^{-}$ decay of the neutron-rich nuclei of ${}^{31}_{14}$Si
could lead to the formation of the nuclei of a stable isotope of phosphorus. In turn, neutron-rich nuclei of ${}^{31}_{14}$Si can be formed 
in a neutron-rich environment through the sequential neutron captures by seed nuclei of stable isotopes of ${}^{28}_{14}$Si, 
${}^{29}_{14}$Si, ${}^{30}_{14}$Si. This can occur during the $s$ process in the massive, intermediate-, and low-mass stars in the final 
stages of their evolution.

To review the phosphorus distribution in cosmos, it is necessary to briefly mention phosphorus presence in the interstellar medium, 
star-forming regions, circumstellar envelopes of late-type  stars, winds and jets of supernovae, novae, post-asymptotic giant branch stars, 
circumstellar disks, Solar System bodies, and stellar atmospheres. In the interstellar medium, the phosphorus was 
found in the form of simple and complex phosphorus-bearing molecules, such as phosphorus monoxide (PO), phosphorus mononitride (PN), phosphine 
(PH$_{3}$), PO$^{+}$, CP, SiP, HCP, and CCP  (see, recent review of \citealt{Fontani2024}). The first detections of the PO and PN molecules 
in the direction of the massive star-forming regions W51 e1/e2 and W3 OH, and a solar-type star-forming region, the dark cloud nebula L1157, 
were reported by \cite{Rivilla2016} and \cite{Lefloch2016}, respectively. PO and PN molecules were detected toward the star-forming 
region AFGL~5142 in the Perseus arm (\citealt{Rivilla2020}). 
 
The P-bearing molecules PN, CP, and HCP were detected in the circumstellar gas of carbon-rich star IRC~+10216 (\citealt{Guelin1990}; 
\citealt{Agundez2007}; \citealt{Milam2008}). \cite{Halfen2008} also detected the carbon monophosphide radical CCP in this 
source. PN and PO molecules were detected in the envelope of the oxygen-rich red supergiant VY~CMa (\citealt{Tenenbaum2007}; 
\citealt{Ziurys2007}; \citealt{Milam2008}). Molecules PN and HCP have been observed in the envelope of the carbon-rich 
protoplanetary nebula CRL~2688, post-AGB star V1610~Cyg (\citealt{Milam2008}). These findings support the hypothesis that a carbon-rich 
environment activates the phosphorus chemistry. PO and PN molecules were detected in the wind of the oxygen-rich AGB star IK~Tau 
(\citealt{DeBeck2013}). Recently, a PN molecule was also detected in the wind of MS-type AGB star RS~Cnc. The discovery of a P-bearing 
molecule PN beyond the Milky Way toward the nearby starburst Galaxy NGC~253 was made by \cite{Haasler2022}. 

Phosphine, PH$_{3}$, has been observed in the atmospheres of Jupiter and Saturn, and P has been identified in meteorites in the mineral 
form and phosphoric acids (see a description of these results and corresponding references in the introduction of \citealt{Rivilla2020}). 
The first detection of the atomic phosphorus in the cometary dust from comet 1P/Halley by the Vega 1 mission in 1986 was recorded by 
\cite{Kissel1987}. The next detection of phosphorus occurred by the NASA Stardust spacecraft during the flyby of comet 81P/Wild in 2004 
(\citealt{Flynn2006}; \citealt{Joswiak2012}). The resulting data were analyzed by \cite{Rotundi2014}. The authors concluded that phosphorus 
was most likely contained within the solid particles. \cite{Altwegg2016} found phosphorus in the gas phase of the comet of 
67P/Churyumov–Gerasimenko (67P/C--G), but without identification of the parent molecule. Finally, \cite{Rivilla2020} reported the discovery 
of the phosphorus monoxide molecule in the gas phase of the comet of 67P/C--G. Later, \cite{Gardneretal2020} published the results of a study 
of the 67P/C--G inner coma, and the detection of the phosphorus in the solid particles. According to the authors this result completes 
the detection of life-necessary CHNOPS-elements in cometary matter. This may provide a strong argument that comets could be a potential
source of these elements on the young Earth. The presence of refractory phosphorus in the HD~100546 star and protoplanetary disk system 
was recently reported by \cite{Kama2025}. To summarize the data on the phosphorus presence in the above-mentioned sources, it can be noted 
that shock-induced chemistry, followed by cosmic-ray-driven photochemistry, can be responsible for the formation of simple and complex 
P-bearing molecular structures.

\begin{figure}
\resizebox{\hsize}{!}{\includegraphics{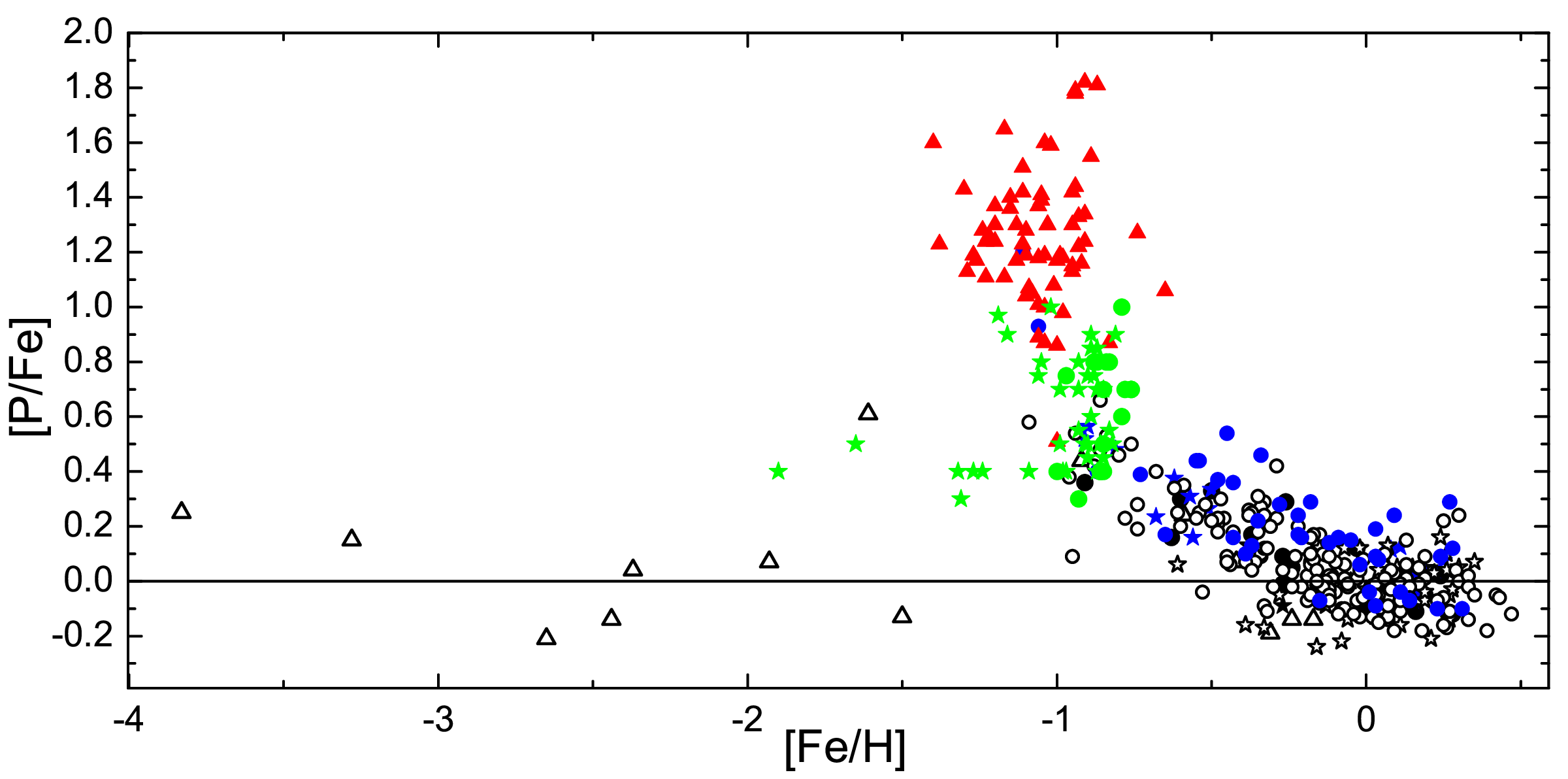}}
\caption{Compiled LTE data on phosphorus abundance in the stars of various metallicities. \cite{Caffau2011} -- filled black circles,
\cite{Roederer2014} -- open triangles, \cite{Caffau2016} -- filled black asterisks, \cite{Maas2019} -- filled blue asterisks, \cite{Maas2022}
-- open circles, \cite{SadakaneNishimura2022} -- open asterisks, \cite{Nandakumar2022} -- filled blue circles, \cite{Brauner2023} -- 
filled red triangles, \cite{Barbuy2025a} -- filled green asterisks, \cite{Barbuy2025b} -- filled green circles.}
\label{Pabund_LTE}
\end{figure}

Separate studies of the phosphorus presence in cosmic objects are devoted to the phosphorus abundance in stars of different ages 
($\sim$ metallicities), because these stars bear the imprints of various cosmic phosphorus suppliers that were active at different stages 
of the evolution of stellar systems. An interesting study of the phosphorus abundance in 82 giants from 24 open clusters and 20 classical 
Cepheids was recently published by \cite{Jian2025}. The authors provide a detailed description of the results on the phosphorus abundance 
determination, starting with the work of \cite{Caffau2011}. In this work, twenty F stars were studied and it was found that the 
relative-to-iron abundance of phosphorus [P/Fe] decreases from +0.4 dex to approximately the solar value as metallicity increases from --1 to 
zero dex. \cite{Caffau2011} also cites several studies on phosphorus abundance determinations conducted before 2011. Dwarf stars 
with a metallicity from --4 to --0.1 dex were studied by \cite{Roederer2014}. The authors confirmed the general trend in the metallicity 
range from --1 to zero dex reported previously by \cite{Caffau2011}, and further found that in the low-metallicity regime ([Fe/H] from 
--4 to --2 dex) the [P/Fe] value remains approximately solar, while in the range of [Fe/H] from --2 to --1 dex the [P/Fe] increases, reaching 
approximately +0.5 dex. Another large-scale study of the phosphorus in the moderately metal-poor stars of globular clusters was initiated 
by \cite{Barbuy2025a} and \cite{Barbuy2025b}. The authors combined their own determinations with results of other works published between 
2011 and 2025 and also showed that phosphorus abundance exhibits a clear sharp increase in the [P/Fe] ratio. Some results of the phosphorus 
abundance determinations under the local thermodynamic equilibrium (LTE) approximation are summarized in Fig. \ref{Pabund_LTE}.

From Fig. \ref{Pabund_LTE} it is evident that the phosphorus abundance distribution over a wide metallicity range appears unusual. 
The nearly vertical [P/Fe] distribution at a metallicity [Fe/H] of approximately --1 dex indicates that alongside phosphorus-normal stars, 
phosphorus-rich stars with a very high ratio [P/Fe] exist in this range. The situation with phosphorus abundance in stars in low-metallicity 
regime is not entirely clear, as data are scarce. To clarify the LTE pattern shown in Fig. \ref{Pabund_LTE} and to draw a more definitive 
conclusion about the phosphorus abundance distribution, it is necessary to reconsider the LTE data using the non-LTE approximation. 
Additional works on the phosphorus abundance in different stars should also be mentioned: \cite{Jacobson2014}, \cite{Masseron2020}, 
\cite{Hinkel2020}, and \cite{Sneden2021}.

From a theoretical perspective, the phosphorus abundance determination in the non-LTE approximation has recently been considered by 
\cite{Takeda2024} (to be discussed below)  and \cite{Aschenbrenner2025}. These authors developed an atomic model for the phosphorus 
ionization stages \ion{P}{ii}, \ion{P}{iii}, and \ion{P}{iv} to study phosphorus lines in the spectra of O and B stars. After analyzing 42 
slowly rotating OB main-sequence stars and B giants in the solar neighborhood, the authors concluded that the present-day cosmic phosphorus 
abundance in this region is $\log \epsilon$ (P/H) + 12 $\equiv$ (P/H) = 5.36 $\pm 0.14$ dex, which is formally slightly lower than the solar 
photospheric abundance of phosphorus (phosphorus abundance in meteorites is 5.43 $\pm 0.03$ dex, \citealt{Lodders2021}).The authors also 
provide a review of studies aimed at determining the phosphorus abundance in hot stars.  

The goal of this work was to develop an atomic model for \ion{P}{i} that could be used to analyze phosphorus lines in the spectra of cool 
dwarf and giants with varying metallicities. This model could help us to study the phosphorus abundance in stars of different ages and 
answer the question of the reality of a strong phosphorus overabundance in P-rich stars.

\section{Phosphorus abundance in the Sun}

Currently, the solar abundance of phosphorus remains uncertain. This is due to the lack of a reliable atomic model of this element. 
The second problem is the quality of the phosphorus lines used for this purpose. Only lines in ultraviolet (UV) and infrared (IR) spectral 
regions can be used for abundance analysis in cool stars. The adopted solar abundance of this element ranges from 
(P/H) = 5.41 $\pm 0.03$ to 5.45 $\pm 0.04$ dex (\citealt{Grevesse1998}; \citealt{Caffau2007}; \citealt{Scott2015}).  
These values were obtained from the analysis of the weak lines (equivalent widths from 1 to 24 m\AA~) in the spectral range from 9500 to 
10800 \AA, which is significantly distorted by telluric lines. It should be noted that the influence of the non-LTE effects on these lines 
has not been properly investigated to date. The main problem is the lack of the necessary atomic data in the literature. 
\cite{Takeda2024} was the first to attempt to create a model of neutral phosphorus. His model is described in the appendix to his paper, 
which is devoted to the study of non-LTE effects affecting the lines of ionized phosphorus in the spectra of B stars. Depending on how the 
collisional rates with hydrogen atoms are taken into account, the model yields a solar phosphorus abundance from 5.31 to 5.40 dex. 
The non-LTE effects lead to an increase in the IR lines. The phosphorus abundance reported by \cite{Aschenbrenner2025} 
(see discussion above) fits well within this range. 

Including a 3D correction to the phosphorus abundance has virtually no effect on the abundance result. This follows from the work 
of \cite{Caffau2007}, which shows that 3D correction to the solar phosphorus abundance is insignificant (see, their Table 4). 
In the next section, we describe our model, which can be used for a non-LTE analysis of \ion{P}{i} UV and IR lines in the spectra of cool stars.

\section{Phosphorus atomic model}

\begin{figure}
\resizebox{\hsize}{!}{\includegraphics{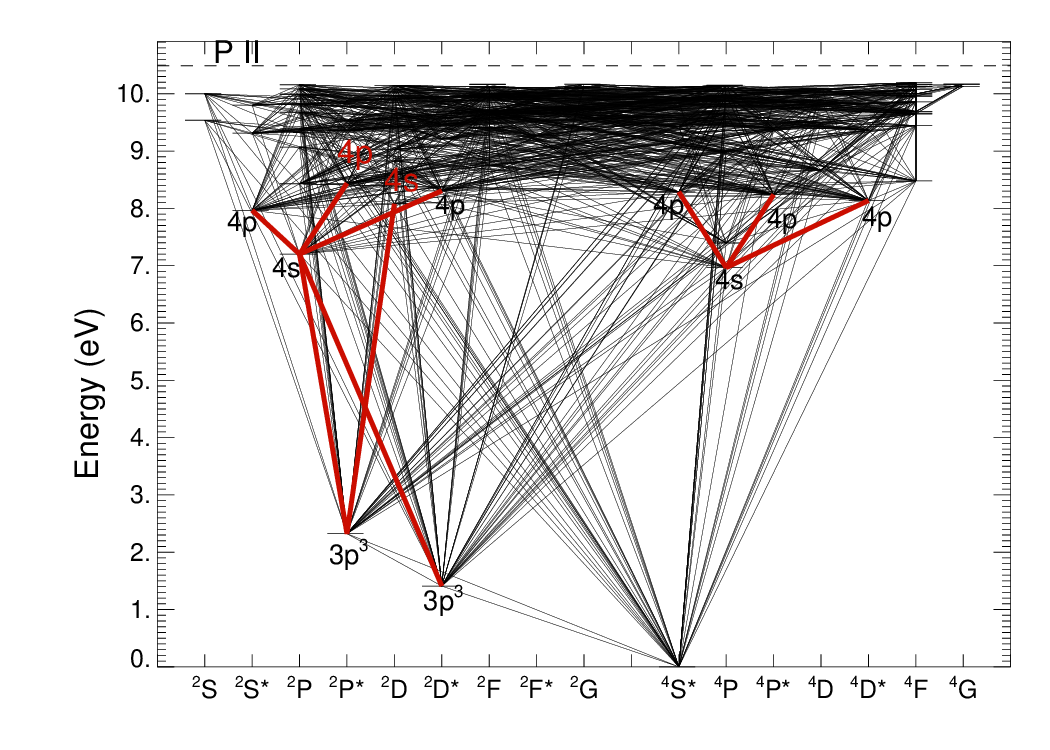}}
\caption{Electronic level and transition diagram for the phosphorus atomic model.}
\label{Grot}
\end{figure}

The model consists of 101 energy levels of \ion{P}{i} and the ground level of \ion{P}{ii}. An additional four levels of \ion{P}{ii} 
with equilibrium populations were added. The fine structure of the levels was not taken into account. The energy level parameters
were taken from \cite{Martin1985}. Ionization energy from the highest level of \ion{P}{i} is 0.29 eV, which corresponds to the  
temperature 3400 \,K. The Grotrian diagram is shown in Fig. \ref{Grot}. A detailed consideration was conducted for 1070 transitions 
between the mentioned levels. We did not take into account transitions producing the lines with wavelengths more than 100000 \AA. 
The oscillator strengths were mainly taken from the Vienna Atomic Line Database (VALD, \citealt{Ryabchikova2015}). In some cases, 
these data were supplemented with data from \cite{Hirata1994} and the webpage of Kurucz (file gfall08oct17.dat). 

Unfortunately, calculations of the phosphorus photoionization cross section were not included in TOPBase (\citealt{Cunto1993}). 
Currently, detailed calculations are only available for the three lowest levels in \cite{Tayal2004}. For the remaining energy levels, 
a hydrogen-like approximation was applied using the effective quantum number, $n^{*}$. For the lower 20 levels, collisional rates with electrons 
were estimated using the calculations from OpenADAS (\citealt{Summers2011}, file ls\#p0.dat). For the remaining allowed transitions, 
as usual, we used the \cite{Regemorter1962} formula, and for the forbidden transitions the collisional strength was set to unity. The collisional 
ionization rate from the ground level was estimated following the method in \cite{Voronov1997}. The collisional ionization rates from  
other levels were calculated according to the \cite{Allen1973} formula with a threshold value of the photoionization cross section. 

As was shown by \cite{Takeda2024}, correctly accounting for the non-LTE effects depends on the methods to consider the phosphorus atom 
collisions with hydrogen. The author used the well-known formula of \cite{Steenbock1984}, which describes the method proposed by 
\cite{Drawin1968}. This method is quite uncertain, even with correction factors, but detailed calculations for phosphorus are lacking; 
therefore, the inelastic collisions with hydrogen in our model were taken into account using the simplified calculation scheme proposed by 
\cite{Belyaev2017a}. According to these authors, this method yields reliable rates in the so-called optimal range from the ionization 
threshold of [--2.7;--1.5] eV. Within the range [--4;--1.3] eV, the accuracy decreases rapidly, and beyond this limits the method fails. 
For the \ion{P}{i} atom these energy regions correspond from 7.79 to 8.99 eV, and from 6.49 to 9.18 eV, respectively. Using this simplified 
method, we calculated tables of collisional rates between the first 31 levels for temperatures between 1000 and 10000 \,K. Rates outside 
the optimal range were evaluated using the \cite{Barklem2017} code, which is based on the free electron model by \cite{Kaulakys1991}. 
Finally, all calculation results were combined. As has been indicated by \cite{Belyaev2017b} and \cite{Belyaev2018}, the difference between 
the results of the simplified model and those of the detailed calculations can reach an order of magnitude. The model of \cite{Kaulakys1991} 
is also unable to guarantee high accuracy. However, they are more realistic than the estimate obtained using the \cite{Steenbock1984} formula, 
which can yield errors of about 5--6 orders of magnitude (\citealt{Barklem2011}). Taking into account the above, it should be noted that 
calculations based on our atomic model of phosphorus cannot claim to be highly accurate. To improve accuracy, detailed quantum mechanical 
calculations are needed.

\section{Spectral line synthesis}

To generate the synthetic spectrum, we used atmosphere models calculated directly using ATLAS9 (\citealt{Castelli2003}) and the opacity 
distribution functions (ODFs) from \cite{Meszaros2012}. These ODFs were constructed using updated absorption lists of molecular lines, 
which is of crucial importance for calculating the atmospheres of cool stars. The population levels of \ion{P}{i} were calculated with the 
help of MULTI code (\citealt{Carlsson1986}) in a modified form proposed by \cite{Korotin1999} that allows the use of ATLAS9 opacities.
Synthetic spectra were obtained using the SynthV code (\citealt{Tsymbal2019}), with the phosphorus line profiles calculated in non-LTE 
using the departure coefficients ($b_{\rm i} = \frac{N_{\rm non-LTE}}{N_{\rm LTE}}$) obtained with MULTI and the profiles of the other 
lines calculated in LTE. Atomic line data for all the lines in vicinity of the line of interest were taken from the VALD database 
(\citealt{Ryabchikova2015}).

\begin{table}
\caption {Phosphorus line parameters and determined phosphorus abundance in the Sun based on the lines used.} 
\label{Lines}
\tiny
\begin{tabular}{lcr|cc|c|c}
\hline
$\lambda$(\AA)& E$_{\rm low}$& $\log~gf$& EW(m\AA)&(P/H)$_{\rm EW}$&(P/H)$_{\rm I}$&(P/H)$_{\rm F}$\\
\hline
 2135.46& 1.40& --1.24&      &     &     &    \\
 2136.18& 1.41& --0.11&      &     &     &    \\
 2149.14& 1.40& --0.36&      &     &     &    \\
 2152.93& 2.32& --0.87&      &     &     &    \\
 2154.07& 2.32& --0.62&      &     &     &    \\
 2154.11& 2.32& --1.32&      &     &     &    \\
 2533.98& 2.32& --1.11&      &     &     &    \\
 2535.61& 2.32& --0.44&      &     &     &    \\
 2553.26& 2.32& --0.86&      &     &     &    \\
 2554.91& 2.32& --1.23&      &     &     &    \\
 9525.74& 6.98& --0.10&  7.7 &5.38 &5.38 &    \\
 9750.74& 6.95& --0.18&  6.3 &5.32 &5.32 &5.36\\
 9790.19& 7.17& --0.69&  0.9 &5.30 &5.32 &    \\
 9796.82& 6.98&   0.27&  6.9 &5.38 &5.34 &5.32\\
 9903.67& 7.17& --0.30&  3.8 &5.39 &5.39 &5.33\\
 9976.68& 6.98& --0.29&  2.9 &5.12 &5.11 &5.22\\
10204.71& 7.21& --0.52&  1.9 &5.31 &5.33 &5.33\\
10511.58& 6.93& --0.13&  7.8 &5.30 &5.34 &5.34\\
10529.52& 6.95&   0.24&  5.3 &5.28 &5.34 &5.32\\
10581.57& 6.98&   0.45&  4.7 &5.34 &5.35 &5.34\\
10596.90& 6.93& --0.21& 10.4 &5.51 &5.39 &5.39\\
10681.40& 6.95& --0.19&  7.8 &5.36 &5.41 &5.40\\
10769.51& 6.95& --1.07&  1.0 &5.31 &5.39 &5.41\\
10813.14& 6.98& --0.41&  4.0 &5.29 &5.34 &5.34\\
11183.24& 7.21&   0.40&      &     &     &    \\
15711.52& 7.17& --0.51&      &     &     &    \\
16482.93& 7.21& --0.29&      &     &     &    \\
\hline                                      
\hline                                      
\end{tabular}                                
\end{table}

Several UV lines and IR lines can be used to determine phosphorus abundance. As was shown by \cite{Roederer2014}, most lines of the three 
multiplets, located in the 2135--2155~\AA~ and 2533--2555~\AA~ranges, are heavily blended. This blending is more pronounced for stars of 
solar metallicity. Only the line 2136~\AA~ is more or less unblended. However, using this line is complicated at solar metallicity. In 
this case, the line is saturated and its profile is virtually insensitive to the change in abundance. This line is suitable for abundance 
determination in metal-poor stars. Another problem arises with IR lines. They are often contaminated by telluric lines. The lines in IR 
region constitute four multiplets (9525--11183~\AA) and one multiplet in the H region (15711--16483~\AA). These lines can be measured in 
spectra of stars with normal metallicity and intermediate metal-poor stars. At a metallicity, [Fe/H], of less than --1.5 dex, these lines 
become undetectable. Table \ref{Lines} lists the parameters of selected \ion{P}{i} lines. The corresponding transitions are highlighted 
in red in Fig. \ref{Grot}. \cite{Berzinsh1997} determined the oscillator strengths as a combination of the experimental data and 
theoretical calculation results. The damping parameters are from \cite{Ryabchikova2015}. For the UV lines, the oscillator strengths 
are from \cite{Wiese1980}. Since these lines are strong, the accurate data of van der Waals constants are needed. These were taken from 
\cite{Roederer2014}. These authors calculated the corresponding damping constants using the method of \cite{Anstee1995}.

The test calculations have shown that the structure of the energy levels in \ion{P}{i} atom is characterized by several factors. Firstly, 
the excited levels are separated from the three lower levels by about 7 eV. Secondly, there are three strong lines of the resonance transitions
$3p^{3}~^{4}S^{*} - 4s~^{4}P$ in the far-UV region (1775--1787~\AA). As a result of the UV pumping, the energy level $4s~^{4}P$ becomes overpopulated.
The level $4s~^{2}P$ in the doublet system is also overpopulated, but to a lesser extent. 

\begin{figure*}
\includegraphics[width= 8.5 cm]{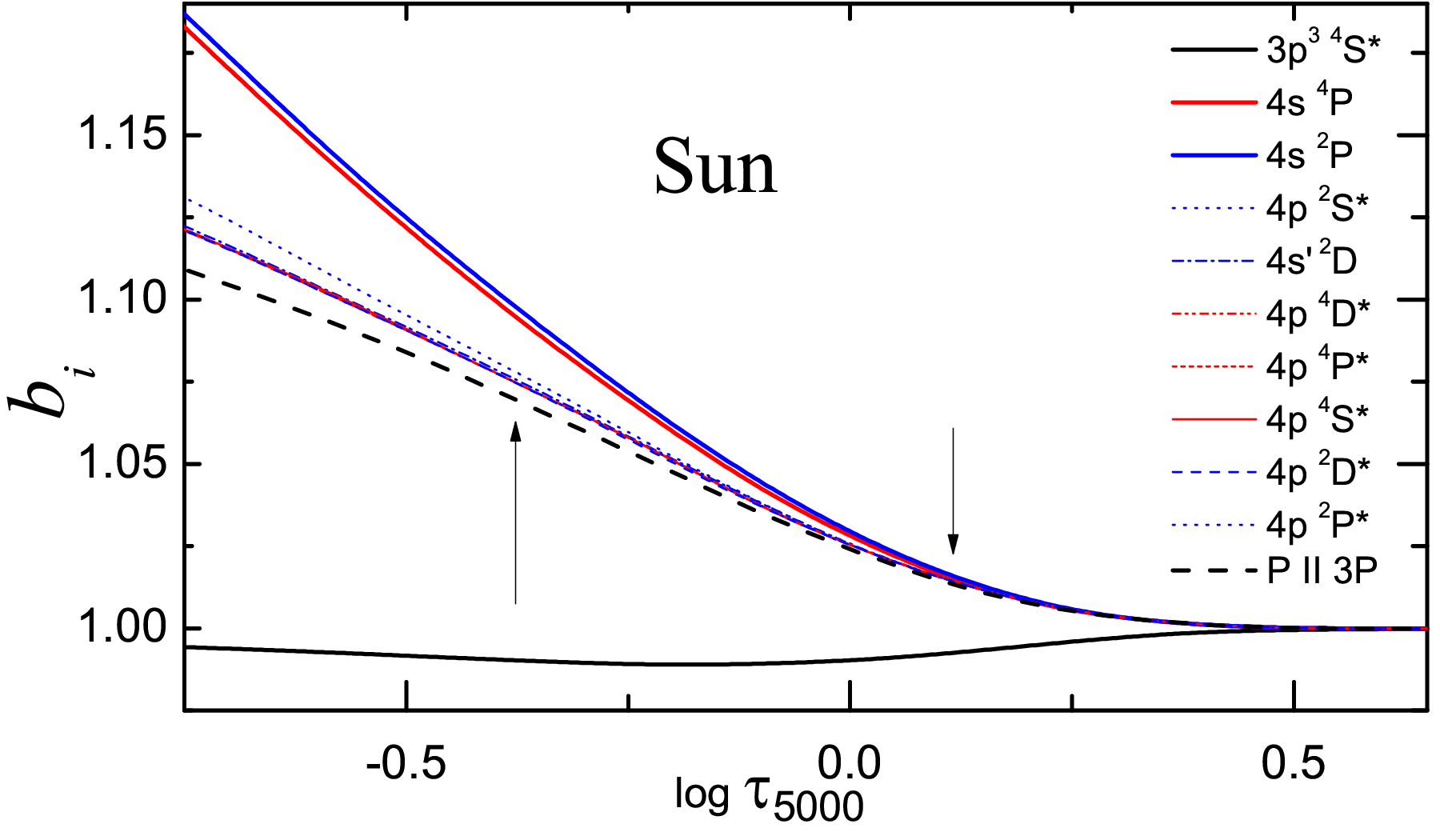}
\includegraphics[width= 8.5 cm]{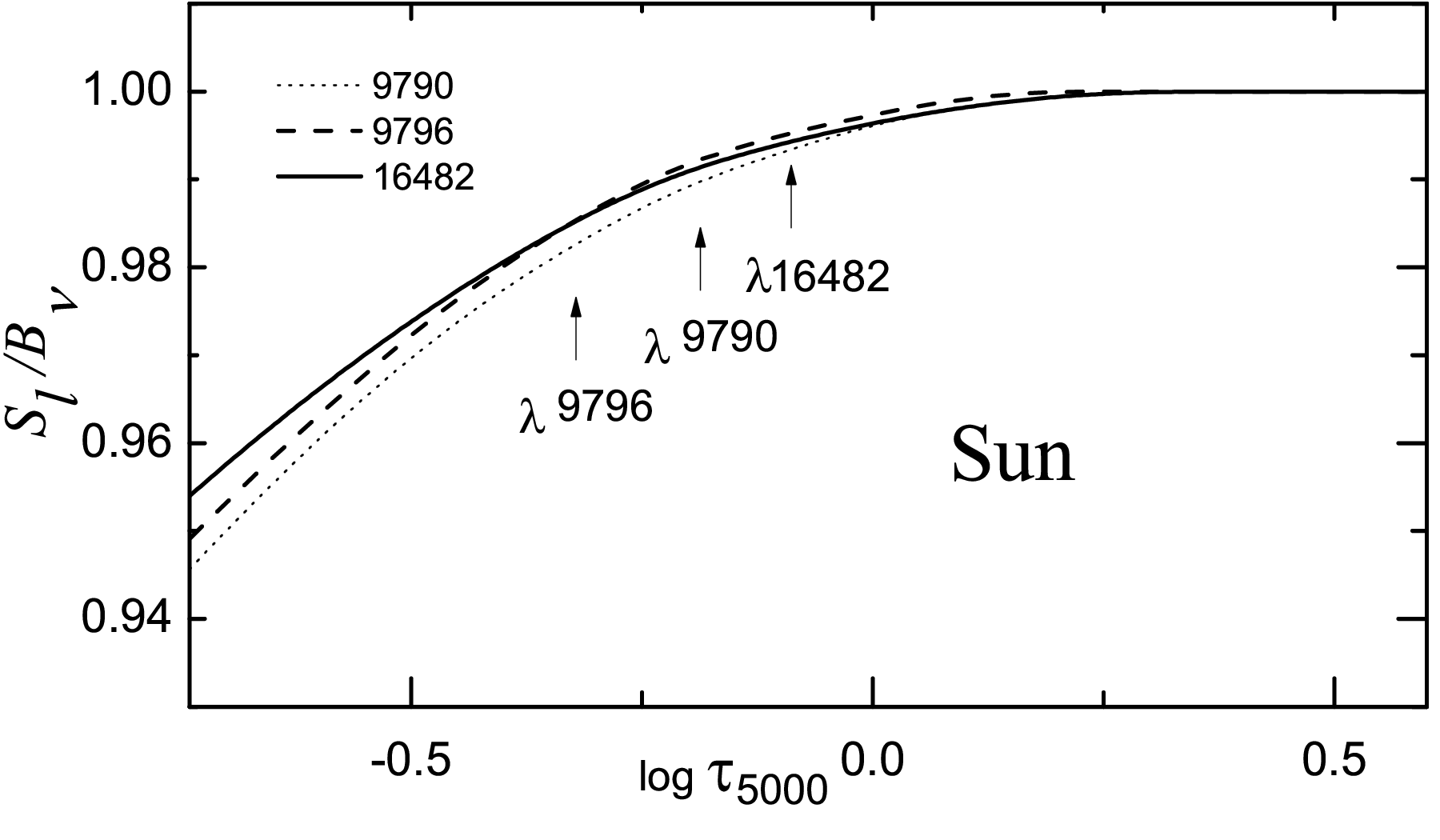}
\caption{Left panel: $b$ factors in the solar atmosphere. Right panel: Change with optical depth in the ratio, $S/B$, between the
source function and Planck function for IR lines. The arrows mark the depth of formation of the cores of some phosphorus lines.}
\label{b_sun}
\end{figure*}

Fig. \ref{b_sun} shows the distributions of the departure coefficients as a function of the optical depth in the solar photosphere 
(to calculate the synthetic spectrum of the Sun, we used the micro- and macroturbulent velocity values of 0.9 and 3.5 km~s$^{-1}$, 
respectively). The ground level of \ion{P}{i} (continuous black line) shows the depopulation, while the ground level of \ion{P}{ii} 
demonstrates overpopulation starting at $\log \tau = 0.5$ (dashed black line). The range of the formation of the discussed IR lines in 
the photosphere is marked by arrows. These lines are formed as a result of transitions from the $4s~^{2}P$ and $4s~^{4}P$ levels. 
The $b$-factor distributions for these lines are shown by the continuous blue and red lines. Clearly, in the region of the IR line 
formation, the population of the lower levels dominates over the population of the upper levels. The right panel of Fig. \ref{b_sun} shows 
that at the line formation depth the source function is smaller than the Planck function. This leads to the line strengthening compared to the 
LTE case. Since the UV lines are very strong, and are effectively formed near the chromosphere, they are excluded from consideration.

\begin{figure*}
\includegraphics[width= 8.5 cm]{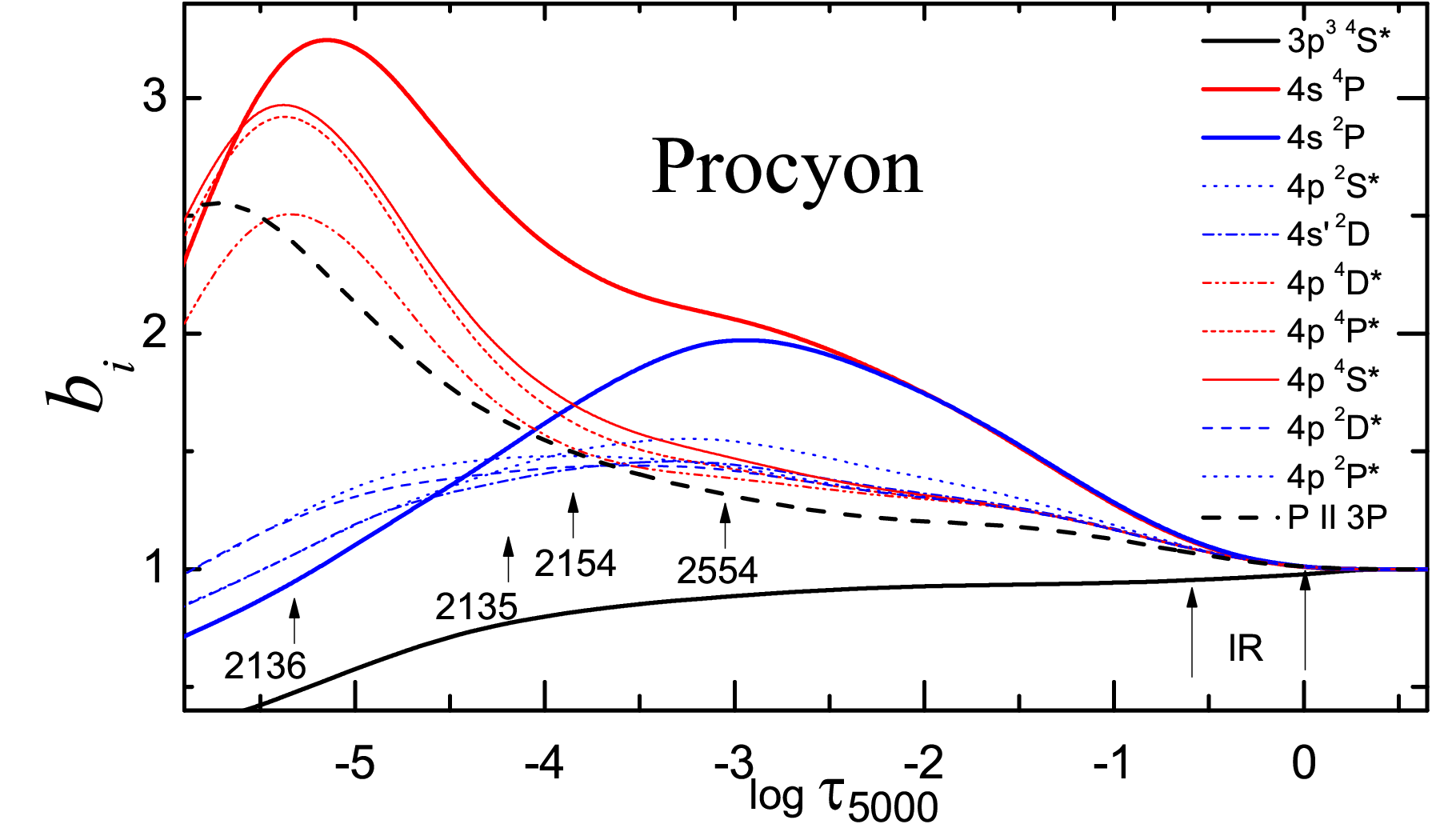}
\includegraphics[width= 8.5 cm]{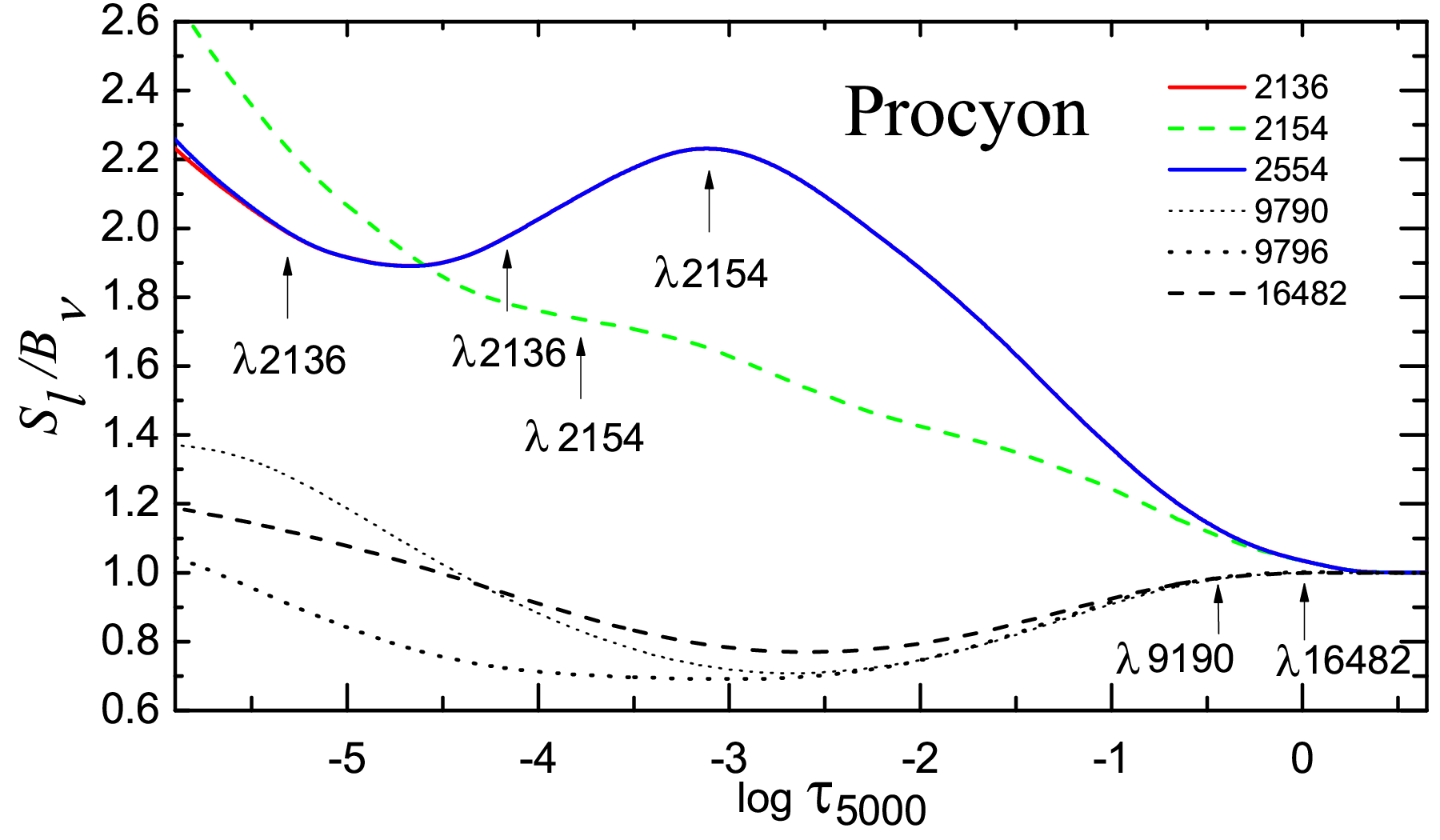}
\caption{Same as Fig. \ref{b_sun} but for Procyon. The locations of the UV line formation are also indicated.}
\label{b_Proc}
\end{figure*}

A similar picture is observed for the IR lines in spectra of the hotter stars. This is evident in the plot for Procyon (Fig. \ref{b_Proc}),
a more detailed discussion is provided below. For such a star, the UV lines are less saturated. They are formed as a result of transitions 
from the levels $3p^{3}~^{2}D^{*}$ and $3p^{3}~^{2}P^{*}$ high in the atmosphere. Their $b$ factors are virtually indistinguishable from 
those for the ground level of \ion{P}{i}. UV lines are formed at different depths and the behavior of their $b$ factors is more complex. 
However, throughout the upper atmosphere, the $b$ factors of the upper levels of these transitions, $4s~^{4}P$, $4s~^{2}P$, and $4s~^{2}D$, 
are larger than those for the lower levels. The ratio of the source function to the Planck function exceeds unity. Consequently, these 
lines are weakened compared to the LTE case.
   
\section{Phosphorus abundance in the solar atmosphere. Our non-LTE result}

\begin{figure*}
\resizebox{\hsize}{!}{\includegraphics{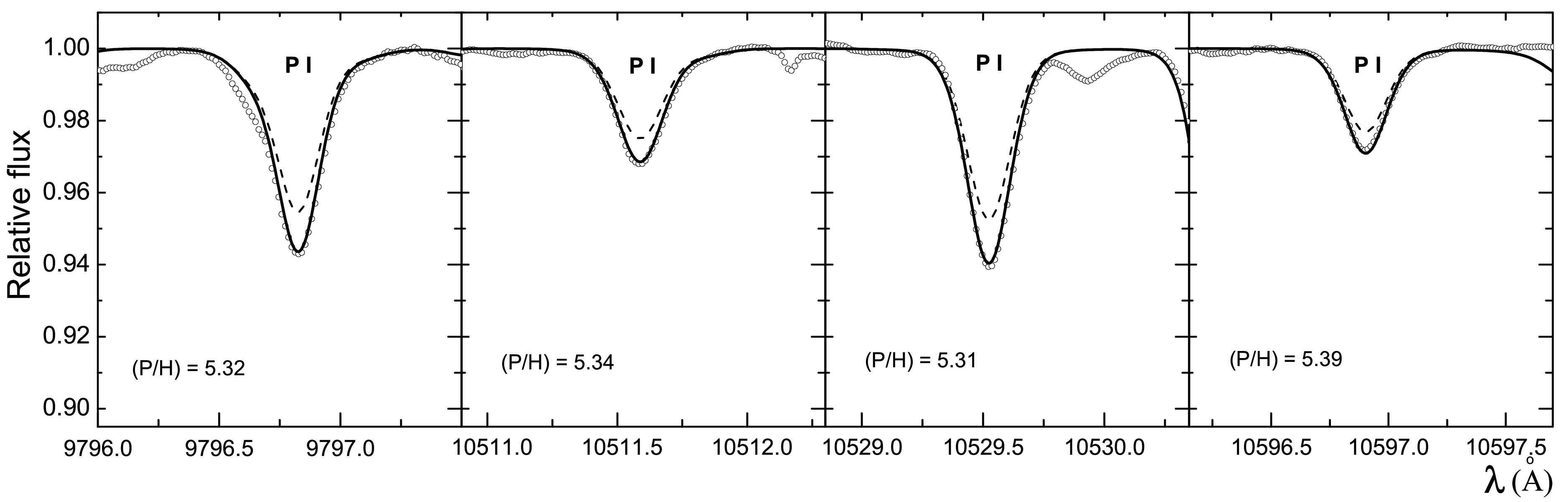}}
\caption{Observed (\citealt{Kurucz1984}, open circles) and synthetic profiles of the phosphorus lines for the Sun (non-LTE profiles -- 
solid line, LTE profiles calculated with the same phosphorus abundance -- dashed line).}
\label{Sun_prof}
\end{figure*}

To determine the phosphorus abundance in the solar atmosphere, we used 14 IR lines, the same as in \cite{Biemont1994}, \cite{Caffau2007}, 
and \cite{Takeda2024}. Parameters of these lines are given in Table \ref{Lines}. Three spectroscopic atlases were used to compare the 
calculated parts of the synthetic spectrum with observed spectra: by intensity, I, \cite{Delbouille1981}, and by flux, F, 
\cite{Kurucz1984}, \cite{Reiners2016}. Moreover, we also used equivalent widths for the lines of interest (\citealt{Biemont1994}) obtained 
from the atlas of \cite{Delbouille1981}. Phosphorus abundance values determined from the two atlases of \cite{Kurucz1984} and 
\cite{Reiners2016} are practically the same. Two lines 9525 and 9790 \AA~ were excluded from analysis, since they are heavily affected by 
the telluric lines. However, these lines are not affected in the \cite{Delbouille1981} spectrum. In Fig. \ref{Sun_prof} we show the fit 
between observed and synthetic profiles for some studied lines.

Table \ref{Lines} lists the phosphorus abundance (P/H) obtained from each line studied. It is seen that the abundance results agree well.
The line 9976 \AA~ shows a clear underabundance. This could a problem caused by the value of its oscillator strength. If we exclude this 
line, the final phosphorus abundance in the Sun is (P/H)=$5.35 \pm 0.06$ dex (equivalent widths). The profile fits give the following results:
(P/H) = $5.36 \pm 0.03$ dex (\citealt{Delbouille1981} spectrum) and (P/H) = $5.35 \pm 0.03$ dex (\citealt{Kurucz1984} and \citealt{Reiners2016}
spectra). Thus, we can conclude that our estimate of the phosphorus abundance in the Sun is (P/H) = $5.35 \pm 0.04$ dex. This value is in 
good agreement with the phosphorus abundance determined by \cite{Aschenbrenner2025} based on an analysis of the \ion{P}{ii} lines for hot 
stars in the solar neighborhood ($5.36 \pm 0.14$ dex), and at the same time slightly lower than the accepted value for meteorites 
($5.41 \pm 0.03$ dex, \citealt{Lodders2021}).

\section{Testing the \ion{P}{i} model using spectra of the stars}

The lines of different multiplets are susceptible to non-LTE effects to varying degrees. Therefore, the optimal check of the model reliability
would be to use the lines of different multiplets. The goal of such a test is to obtain similar phosphorus abundance values for all lines 
studied, which would indicate the correctness of the atomic model. As to \ion{P}{i} lines, such a check is complicated by the fact
that in the case in which UV lines are not yet saturated and suitable for analysis, the IR lines are very weak, often distorted by the
telluric lines, and cannot be used in analysis. On contrary, when IR lines are reliably detected, the UV lines are saturated and insensitive
to the abundance variation. An optimal case is observed for the stars with a solar-like metallicity and an effective temperature in the range 
of 6000 -- 6500 \,K. Naturally, the high-resolution spectra in UV and IR range must be available. To follow this recipe, we selected the
spectra of Procyon obtained with the Space Telescope Imaging Spectrograph (STIS; \citealt{Kimble1998}; \citealt{Woodgate1998}) on board 
the Hubble Space Telescope (HST) and the high-resolution spectropolarimeter NARVAL  (\citealt{Auriere2003}). The spectra were taken
from the Barbara A. Mikulski Archive for Space Telescopes (MAST) and archive PolarBase  (\citealt{Petit2014}). The following parameters of
Procyon were used: T$_{\rm eff} = 6582$ \,K, $\log~g = 3.98$ dex,  V$_{\rm t} =1.7$ km~s$^{-1}$, and [Fe/H] = --0.02 dex (\citealt{Soubiran2024}). 
From the IR region, the only 9796 and 9903 \AA~ lines are available, which gives a phosphorus abundance (P/H) = $5.35 \pm 0.05$ dex. 
The LTE phosphorus abundance determined from these lines is 5.60 dex, giving a non-LTE correction of about 0.25 dex. Line 2136~\AA~ is 
virtually insensitive to non-LTE effects and yields (P/H) = $5.33 \pm 0.10$ dex. The declared error is due to uncertainty in the 
continuum placement and profile fitting. Thus, a good agreement can be noted between the obtained values of the non-LTE phosphorus abundance.

\begin{figure*}
\resizebox{\hsize}{!}{\includegraphics{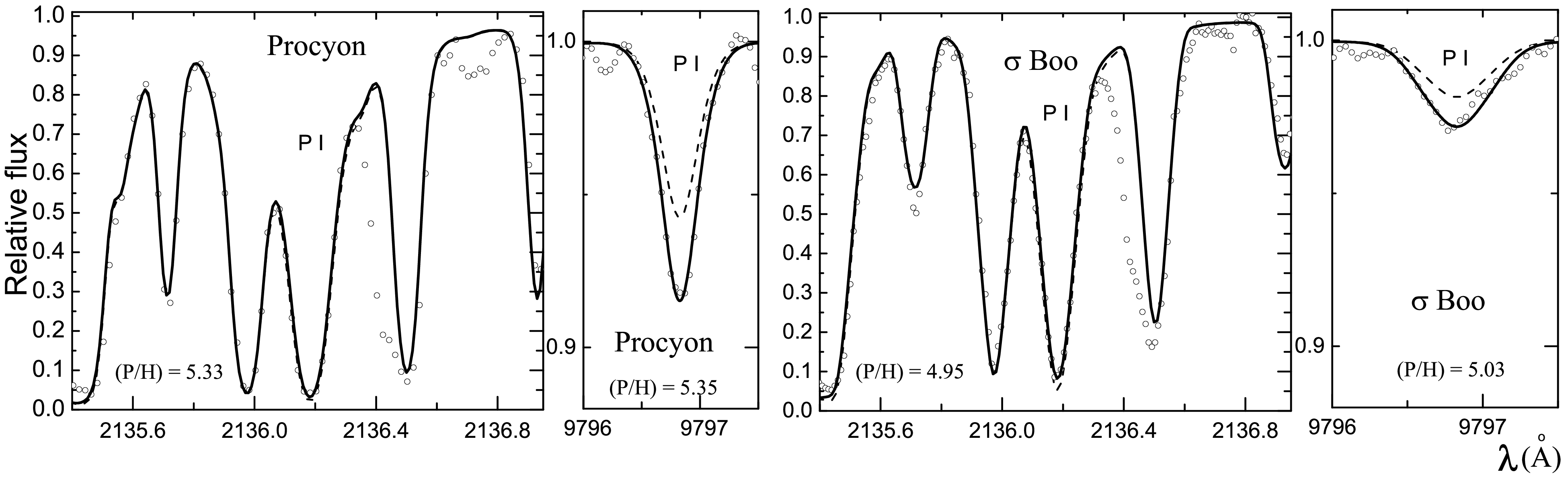}}
\caption{Observed (open circles)  and synthetic profiles of the phosphorus lines in the spectra of Procyon and $\sigma$~Boo. 
The non-LTE profiles are represented by the continuous line, and the LTE profiles calculated with the same phosphorus abundance 
by the dashed line.}
\label{Prof_stars}
\end{figure*}

The second star that meets the above criteria is $\sigma$~Boo. We used the following parameters of this star: T$_{\rm eff} = 6756$ \,K, 
$\log~g = 4.28$ dex, V$_{\rm t} = 1.9$ km~s$^{-1}$, and [Fe/H] = --0.36 dex (\citealt{Luck2018}). Spectral material was taken from MAST 
(STIS HST) and PolarBase (ESPaDOnS). The line 2136~\AA\ yields the phosphorus abundance (P/H) = $4.99 \pm 0.10$ dex; the non-LTE correction 
is +0.08 dex. At the same time, two IR lines 9750 and 9796~\AA\ yield an abundance (P/H) = $4.95 \pm 0.05$ dex; the non-LTE correction is 
about --0.23 dex. In other words, the difference between the LTE phosphorus abundance values obtained from UV and IR lines exceeds 0.3 dex, 
while it is only 0.04 dex after taking into account the non-LTE corrections. Fundamental parameters for this star are less reliable than 
for Procyon, and therefore we used another set of parameters provided by \cite{Garcia2021}: T$_{\rm eff} = 6875$ \,K, $\log~g = 4.62$ dex, 
V$_{\rm t} = 1.9$ km~s$^{-1}$, and [Fe/H] = --0.48 dex. In this case, the line 2136~\AA\ gives the phosphorus abundance 
(P/H) $= 4.95 \pm 0.10$ dex, while the IR lines produce an abundance (P/H) $= 5.03 \pm 0.05$ dex. We see that the difference between abundances
from UV and IR lines increases to 0.08 dex, but it is still four times smaller than in the case of LTE analysis. Fig. \ref{Prof_stars} shows
the result of fitting the observed and synthetic line profiles for these two studied stars. The good agreement between the phosphorus 
abundances obtained from different lines, which have non-LTE corrections that differ not only in magnitude but also in sign, gives hope 
that the developed model is sufficiently reliable and can be used in the non-LTE spectroscopic analysis of other stars. 

\section{Grid of the non-LTE corrections}

The most precise way to determine the non-LTE abundance is the direct synthesis of profiles of selected lines and their subsequent 
comparison with the corresponding profiles in the observed spectrum. However, to roughly estimate the influence of the non-LTE effects on  
a given abundance, it is sufficient to use the LTE abundances and non-LTE corrections calculated for the grid of stellar parameters. 
It should be noted that this method has accuracy limitations and does not take into account the influence of non-LTE effects on line 
profiles, since equivalent widths are used. 
The studied lines can be divided into two groups depending on the size of the non-LTE correction. Lines of a given multiplet with close
intensities should give close non-LTE corrections. Therefore we selected 12 such lines: 2135~\AA\ (2149~\AA), 2136~\AA, 2154~\AA\ 
(2152~\AA), 2534~\AA\ (2535~\AA, 2553~\AA, 2554~\AA), 9525~\AA, 9750~\AA\ (9976~\AA), 9796~\AA, 9903~\AA\ (9790~\AA, 10204~\AA), 
10539~\AA\ (10581~\AA, 10596~\AA), 10813~\AA\ (10511~\AA, 10681~\AA, 10769~\AA), 11183~\AA, and 16482~\AA\ (15711~\AA). Here, in 
brackets, the lines with similar non-LTE corrections are listed. To create the grid of corrections, we used the atmosphere models
calculated with the ATLAS9 code and ODFs from \cite{Meszaros2012}. The calculations were carried out for the parameter ranges: T$_{\rm eff}$ 
from  4000 to 6750 \,K, step 250 \,K; $\log~g$ from 1 to 5 dex, step 1 dex; and V$_{\rm t} = 2$ km~s$^{-1}$, [Fe/H] from --3 to +0.5 dex, 
step 0.5 dex. The non-LTE corrections were calculated for LTE phosphorus abundance ratios [P/Fe] = --0.4, 0.0, and +0.4 dex. The LTE solar
phosphorus abundance is (P/H) = 5.43 dex. For metal-poor stars, the non-LTE corrections are given if the UV lines have equivalent widths
larger than 5 m\AA, while for IR lines the equivalent widths exceed 1 m\AA. It should be noted that UV lines in the spectra of
cool giants suffer from the specific structure of their upper atmospheres. Since we have no information about this structure, 
the non-LTE corrections for UV lines for such stars were excluded from consideration as uncertain. Generally, the non-LTE
corrections for the giant stars ($\log~g = 1$ dex) should be used with caution due to the unaccounted influence of the structure
of the upper atmospheres, as well as the uncertainty in the calculated collisional rates with hydrogen.

\section{Discussion}

We developed an atomic model of \ion{P}{i} for analyzing phosphorus lines in the spectra of cool stars with different metallicities. This 
model was used to calculate the grid of non-LTE corrections for a fairly wide range of effective temperatures and gravities. Phosphorus 
lines from UV and IR spectral regions were analyzed. The distribution of the LTE phosphorus abundance as a function of metallicity 
(Fig. \ref{Pabund_LTE}) exhibits a complex structure. Data on the LTE phosphorus abundance are scarce in the low-metallicity region, while 
the higher-metallicity region is well populated by the stars studied. Our primary interest is in metal-poor stars and stars with a high 
relative-to-iron phosphorus abundance (P-rich stars). 

\begin{figure}
\resizebox{\hsize}{!}{\includegraphics{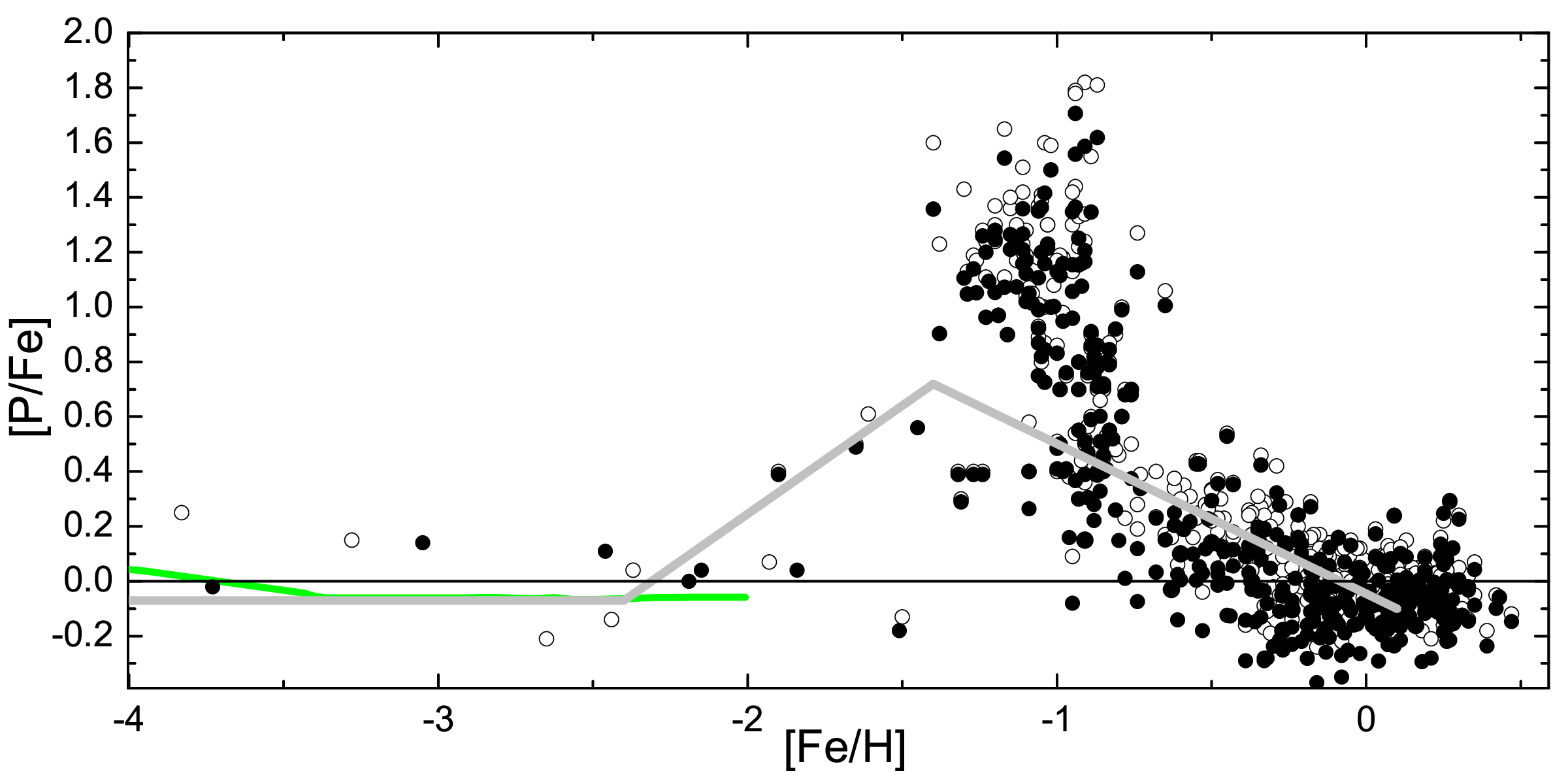}}
\caption{Compilative LTE data on phosphorus abundance from Fig. \ref{Pabund_LTE} (open circles) with our non-LTE corrections (filled
circles). The teoretical prediction given in \cite{BekkiTsujimoto2024} is shown by the gray line. For the low-metallicity region, 
the theoretical prediction from \cite{Prantzos2018} is shown by the green line.}
\label{Pabund_NLTE}
\end{figure}

We corrected the LTE phosphorus abundance data presented in Fig. \ref{Pabund_LTE} using our calculated grid of non-LTE corrections. 
The results are shown in Fig. \ref{Pabund_NLTE}. In the case of \cite{Roederer2014}, the data for 14 stars were corrected 
by direct non-LTE line profiles reanalysis. For this, we used the same spectral material obtained with STIS HST (MAST archive). Only 
one line, 2136 \AA, was used for analysis, since most UV lines are heavily blended. In recent years, the parameters of these 14 stars 
have been redefined. Four stars were analyzed by \cite{Mittal2025} as part of their study of metal-poor star abundances. Seven stars 
can be found in catalog of atmospheric parameters of FGK stars (\citealt{Soubiran2022}). HD~140283 was studied in detail by \cite{Creevey2015}. 
Two stars were studied by \cite{Maas2022}, with their phosphorus abundance determined using the 10529~\AA\ line. The parameters of the 
program stars are listed in Table \ref{StarsRoederer}. These parameters were used to recalculate the phosphorus abundance (P/H) and [P/Fe] with 
the solar abundance value (P/H) = 5.35 dex. It should be noted that the change in the phosphorus abundance [P/Fe] occurs not only due to 
non-LTE consideration, but also due to the change in [Fe/H], which depends on atmosphere parameters of program stars. 

Inspecting Fig. \ref{Pabund_NLTE}, one can note that the non-LTE corrections do not qualitatively alter the overall phosphorus abundance 
distribution over a wide metallicity range. At a metallicity [Fe/H] lower than --1 dex, the non-LTE corrected abundances are now distributed
more compactly around [P/Fe] = 0 dex. The cluster of stars around [Fe/H] $\approx 0$ dex exhibits a slightly larger phosphorus deficiency 
than that seen in the LTE plot. What is interesting to note is that the non-LTE corrections did not remove the remarkable phosphorus 
overabundance for the stars with a metallicity of around --1 dex. The existence of such P-rich stars is real. Future non-LTE studies
should clarify the mechanisms of the phosphorus production in the Universe and the origin of P-rich stars.

\cite{Medhi2025} discussed a “silicon--phosphorus” nuclear cycle to explain the phosphorus production. The authors believe that the 
following chain plays a certain role: ${}^{28}_{14}$Si(${}^{1}_{1}$H,$\gamma$)${}^{29}_{15}$P(${}^{1}_{1}$H,$\gamma$)${}^{30}_{16}$S($e^{+} \nu$)
${}^{30}_{15}$P(${}^{1}_{1}$H,$\gamma$)${}^{31}_{16}$S($e^{+} \nu$)${}^{31}_{15}$P(${}^{1}_{1}$H,${}^{4}_{2}$He)${}^{28}_{14}$Si. 
These reactions might be expected in the massive stars, but these authors propose an alternative scenario based on the possible formation 
of these isotopes even in Sun-like stars under specific temperature and density conditions.

\cite{Masseron2020} 
discussed several possibilities that could explain the existence of P-rich stars. These stars could acquire additional phosphorus from their 
unseen companions, similar to CH-stars -- members of binary star systems. Unfortunately, most observed P-rich stars do not exhibit noticeable 
changes in radial velocity. These stars could be the post-AGB stars. However, known P-rich stars do not meet 
the criteria for post-AGB stars. Another possibility is that P-rich stars formed from gas contaminated by the nearby phosphorus sources. 
For example, these sources could be low-mass AGB stars or massive stars. In the former case, nuclei of a stable phosphorus isotope can be 
produced after the neutron captures by ${}^{30}_{14}$Si nuclei. It is believed  that this scenario is responsible for the emergency of 
the aforementioned phenomenon of CH-stars. Since P-rich stars exhibit a different abundance pattern compared to CH-stars, the authors ruled 
out this hypothesis, and support the hypothesis of core-collapse supernovae being the main source of phosphorus in the interstellar medium. 
This is true, but given the masses of the progenitors of these stars, they should have effectively contributed the phosphorus to the 
interstellar medium in the early stages of the Galaxy evolution. This source was apparently insufficient in a later era when the metallicity 
[Fe/H] was $\approx -1 $ dex. A similar conclusion, but for AGB stars, was reached by \cite{KarakasLugaro2016}. \cite{LeungNomoto2018} 
do not consider the carbon-oxygen white dwarf explosions in Type Ia supernovae to be a significant source of the phosphorus. The same problem 
arises with the pair-instability supernovae (thermonuclear explosions of very massive stars). According to \cite{Kozyreva2014} and 
\cite{Takahashi2018}, this type of supernova is characterized by a reduced production of nuclei of odd-Z chemical elements, in particular
phosphorus.

\cite{Brauner2023} consider nonstandard phosphorus production to be a result of the nonthermal nucleosynthesis (\citealt{Goriely2022}).
This is based on the supposition about high-energy reactions involving the $\alpha$ particles and protons. The main drawback of this 
hypothesis is the lack of a specific mechanism for accelerating charge particles to the high energies (a few mega-electronvolts).
 
As can be seen, the situation with phosphorus production during the Galaxy evolution and formation of the P-rich stars remains unclear.
This situation appears to be changing following the publication of \cite{BekkiTsujimoto2024}. Specifically, the authors showed that the 
overall phosphorus abundance distribution on metallicity can be reproduced well by considering two sources of phosphorus production: 
core-collapse supernovae explosions and oxygen-neon novae explosions (with carbon-oxygen white dwarfs of about 1.25 M$_{\odot}$ and less 
massive white dwarfs). These sources are effective on different timescales. The frequency of the oxygen-neon novae outbursts in our Galaxy 
was estimated in \cite{Kemp2022} and \cite{De2021}. According to Fig. 1 of \cite{BekkiTsujimoto2024}, the increase in the relative-to-iron
phosphorus abundance around [Fe/H] $\approx -2.4$ dex and the maximum of [P/Fe] observed at [Fe/H] $\approx -1.5$ dex are due to the 
effective phosphorus contribution to the interstellar medium from outbursts of oxygen-neon novae (in fact, oxygen-neon-magnesium, ONeMg, 
novae consisting of ONeMg white dwarfs), which account for about 30\% of the classified novae events. For example, the detection of 
phosphorus (IR observations) in the envelope of the classical nova V1974~Cyg was reported by \cite{WagnerDePoy1996}. Spectra of the white 
dwarf in the cataclysmic binary VW~Hyi, obtained with the Hubble Space Telescope (Goddard High-Resolution Spectrograph). These spectra, 
exposed a month after the end of a normal dwarf nova outburst, revealed a phosphorus abundance approximately three orders of magnitude 
higher than in the Sun (\citealt{Sion1997}). A large enhancement of phosphorus in quasar broad absorption line outflow regions was noted 
in \cite{Junkkarinen1995}, \cite{Shields1996}, \cite{Turnshek1996}, \cite{Junkkarinen1997}, and \cite{Hamann1998}. This phenomenon is 
believed to be caused by explosions of ONeMg novae. Interestingly, this type of novae, as efficient phosphorus producers, may have been 
the first to be studied by \cite{Wanajo1999}. A high phosphorus abundance is achieved at the peak temperature above $3 \times 10^{8}$ \,K 
in reactions between protons and ${}^{30}_{15}$P nuclei in the silicon--phosphorus cycle.

Finally, it should emphasized again that after the non-LTE corrections, the phosphorus abundance distribution became more compact in the 
low-metallicity range with a mean value of [P/Fe] $\approx 0$ dex. In this domain, the theoretical prediction of \cite{Prantzos2018}, 
which includes massive stars with a combined effect of metallicity, rotation, and mass loss, describes the abundance distribution reasonable 
well. For a more detailed study of the phosphorus abundance distribution in this metallicity region, additional UV spectroscopic 
observations are needed.

\section{Conclusion}

The goal of this work is to develop an atomic model for \ion{P}{i} that can be used to analyze phosphorus lines in the spectra of cool 
dwarf and giants with varying metallicities in the non-LTE approximation. Our grid of non-LTE corrections, as well as direct line profile 
synthesis, was used to refine the literature data on phosphorus abundance in metal-poor, intermediate-deficient, and solar-metallicity 
stars. This sample also includes phosphorus-rich stars. Non-LTE corrections do not qualitatively alter the overall phosphorus abundance 
distribution over a wide metallicity range, and do not change the characteristic pattern of phosphorus-rich stars. After corrections, 
the phosphorus abundance distribution became more compact in the low-metallicity range. Overall, the observed phosphorus abundance 
distribution can be described by the combined effect of phosphorus production in massive rotating stars, Type II supernovae explosions, 
and oxygen-neon-magnesium novae.

\begin{table}
\caption {Redetermined non-LTE phosphorus abundance in the stars from \cite{Roederer2014}.} 
\label{StarsRoederer}
\begin{tabular}{lcccrc}
\hline
Star& T$_{\rm eff}$, K&$\log~g$&[Fe/H]&[P/Fe]&Rem\\
\hline
BD+44  493&5463 &3.20&--3.73 &--0.02 &1\\
G64--12   &6541 &4.26&--3.05 &  0.14 &1\\
HD107113  &6424 &4.10&--0.53 &  0.03 &2\\
HD108317  &5365 &2.84&--2.19 &  0.00 &1\\
HD128279  &5331 &3.12&--2.15 &  0.04 &1\\
HD140283  &5642 &3.65&--2.46 &  0.11 &3\\
HD155646  &6168 &3.74&--0.14 &--0.19 &2\\
HD160617  &5963 &3.73&--1.84 &  0.04 &2\\
HD16220   &6199 &3.93&--0.28 &--0.11 &2\\
HD211998  &5243 &3.44&--1.51 &--0.18 &2\\
HD2454    &6508 &4.06&--0.24 &--0.07 &4\\
HD43318   &6179 &3.81&--0.24 &--0.09 &4\\
HD76932   &5871 &4.11&--0.88 &  0.28 &2\\
HD94028   &5967 &4.33&--1.45 &  0.56 &2\\
\hline                                      
\hline                                      
\end{tabular}                                
\tablebib{Stellar parameters are from (1)~\citet{Mittal2025};
(2) \citet{Soubiran2022}; (3) \citet{Creevey2015}; (4) \citet{Maas2022}.}
\end{table}

\section{Data availability}
A grid of non-LTE corrections as function of the stellar parameters is 
available on \href{https://doi.org/10.5281/zenodo.18860904}{https://doi.org/10.5281/zenodo.18860904} 
The data are presented both in the graphical form and in the form of machine-readable table.

\begin{acknowledgements}

We are deeply grateful to our referee for carefully reading our article and providing helpful comments
that improved this work.

\end{acknowledgements}

\bibliographystyle{aa}
\bibliography{P_I}

\begin{appendix}
\onecolumn
\section{Grid of the non-LTE corrections.}

\begin{figure*}
\includegraphics[scale=0.5]{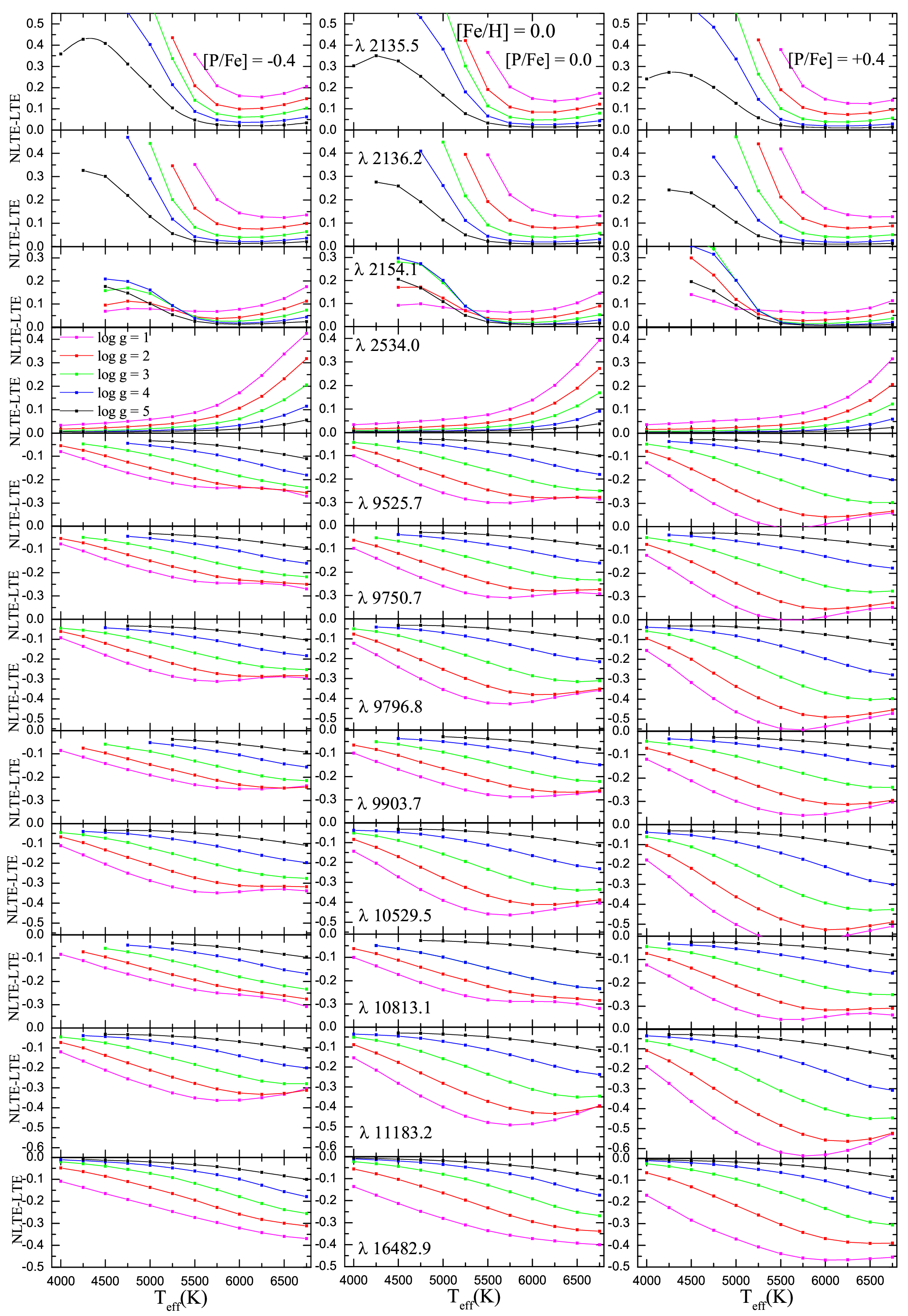}
\caption{Non-LTE corrections for metallicity [Fe/H]=+0.5}
\label{dnlte_p05}
\end{figure*}

\begin{figure*}
\includegraphics[scale=0.5]{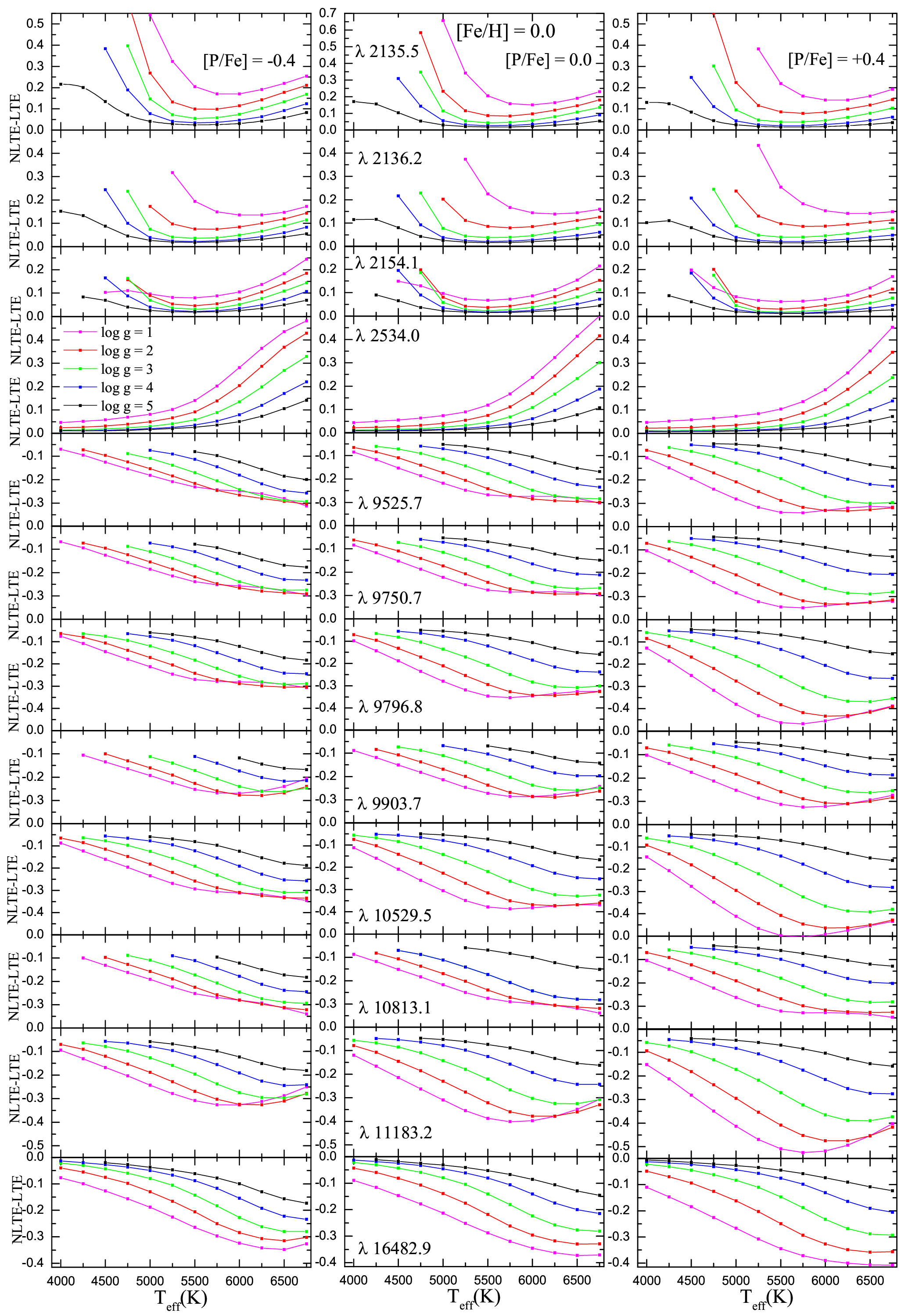}
\caption{Non-LTE corrections for metallicity [Fe/H]=0.0}
\label{dnlte_p00}
\end{figure*}

\begin{figure*}
\includegraphics[scale=0.5]{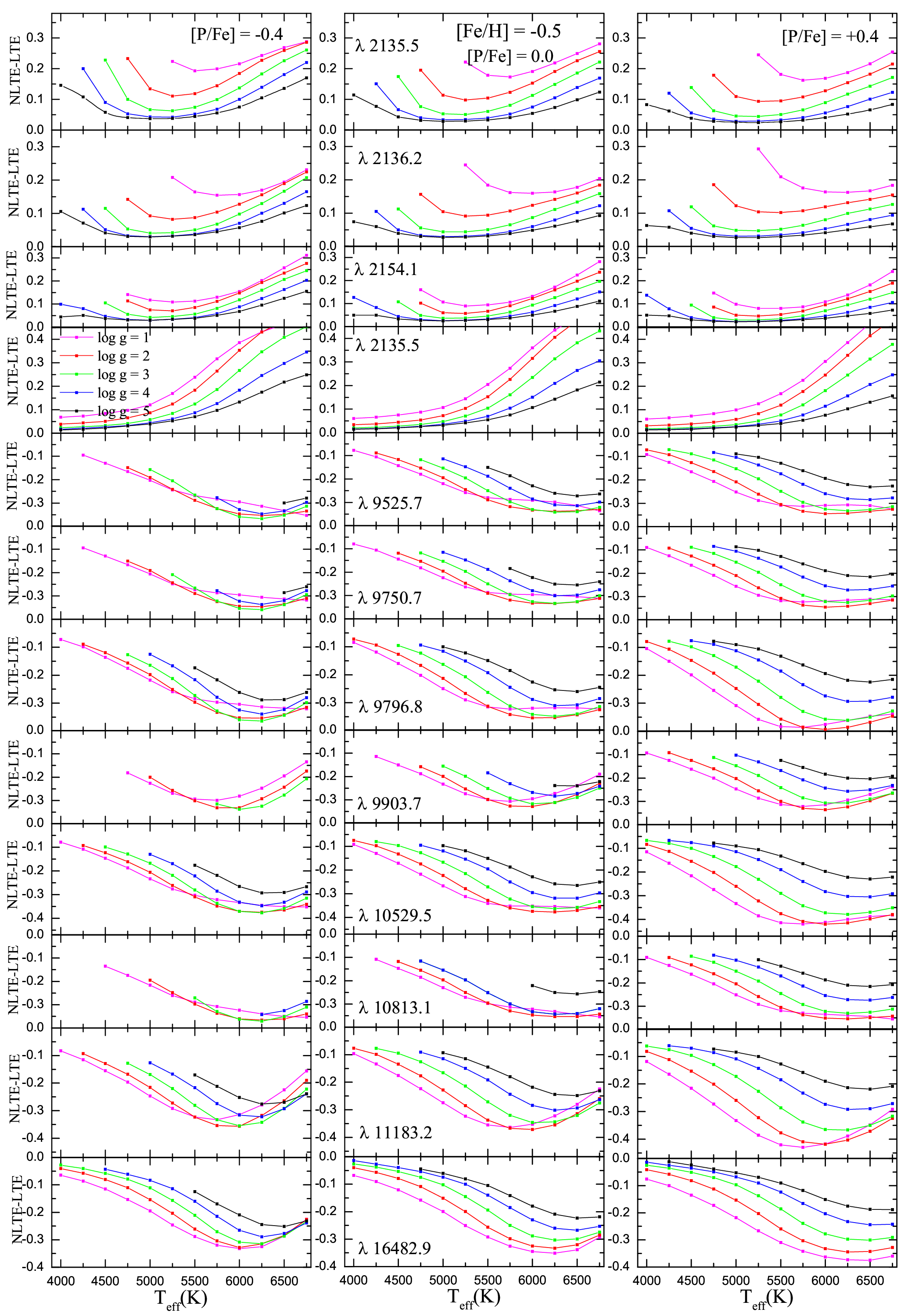}
\caption{Non-LTE corrections for metallicity [Fe/H]=--0.5}
\label{dnlte_m05}
\end{figure*}

\begin{figure*}
\includegraphics[scale=0.5]{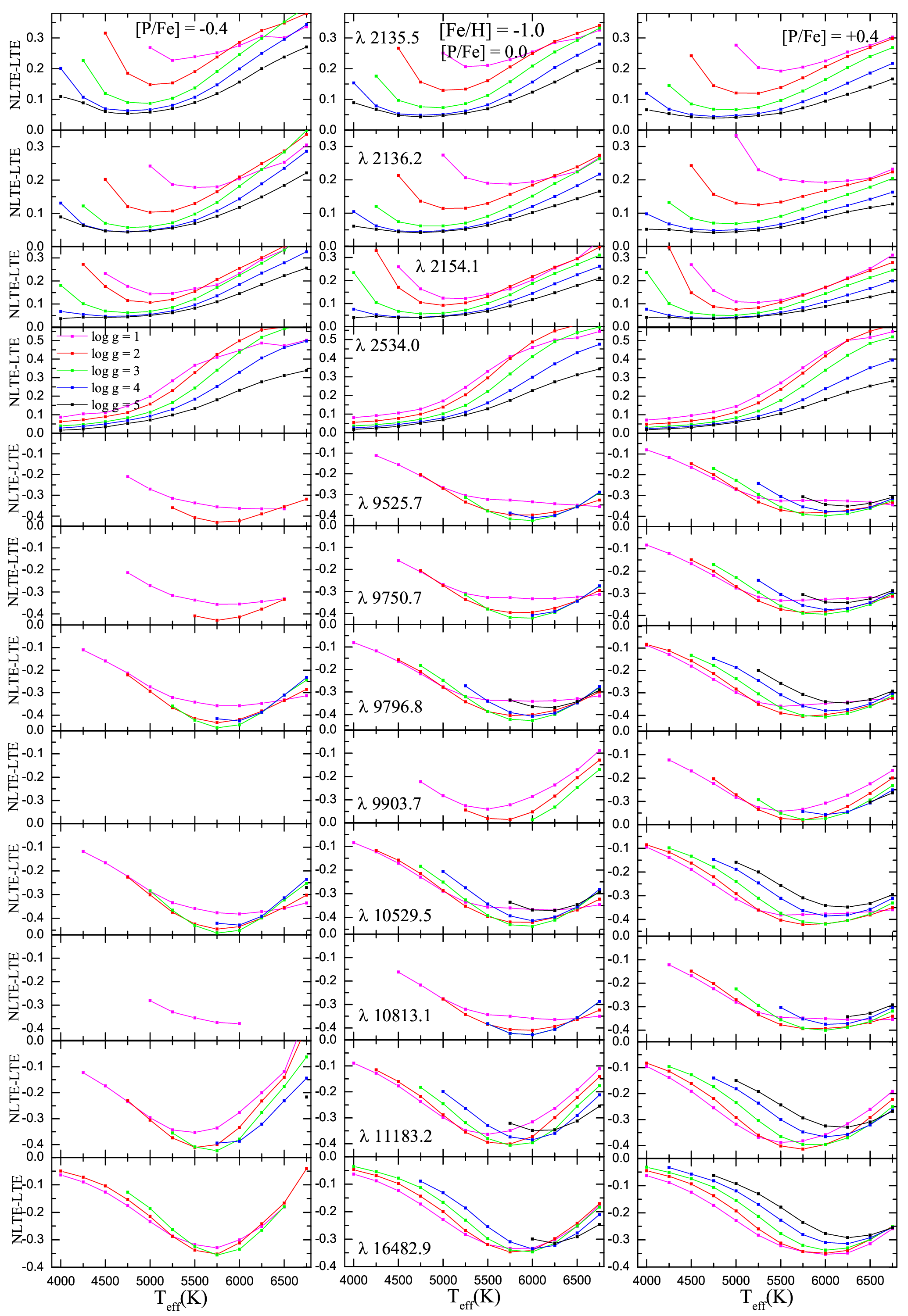}
\caption{Non-LTE corrections for metallicity [Fe/H]=--1.0}
\label{dnlte_m10}
\end{figure*}

\begin{figure*}
\includegraphics[scale=0.5]{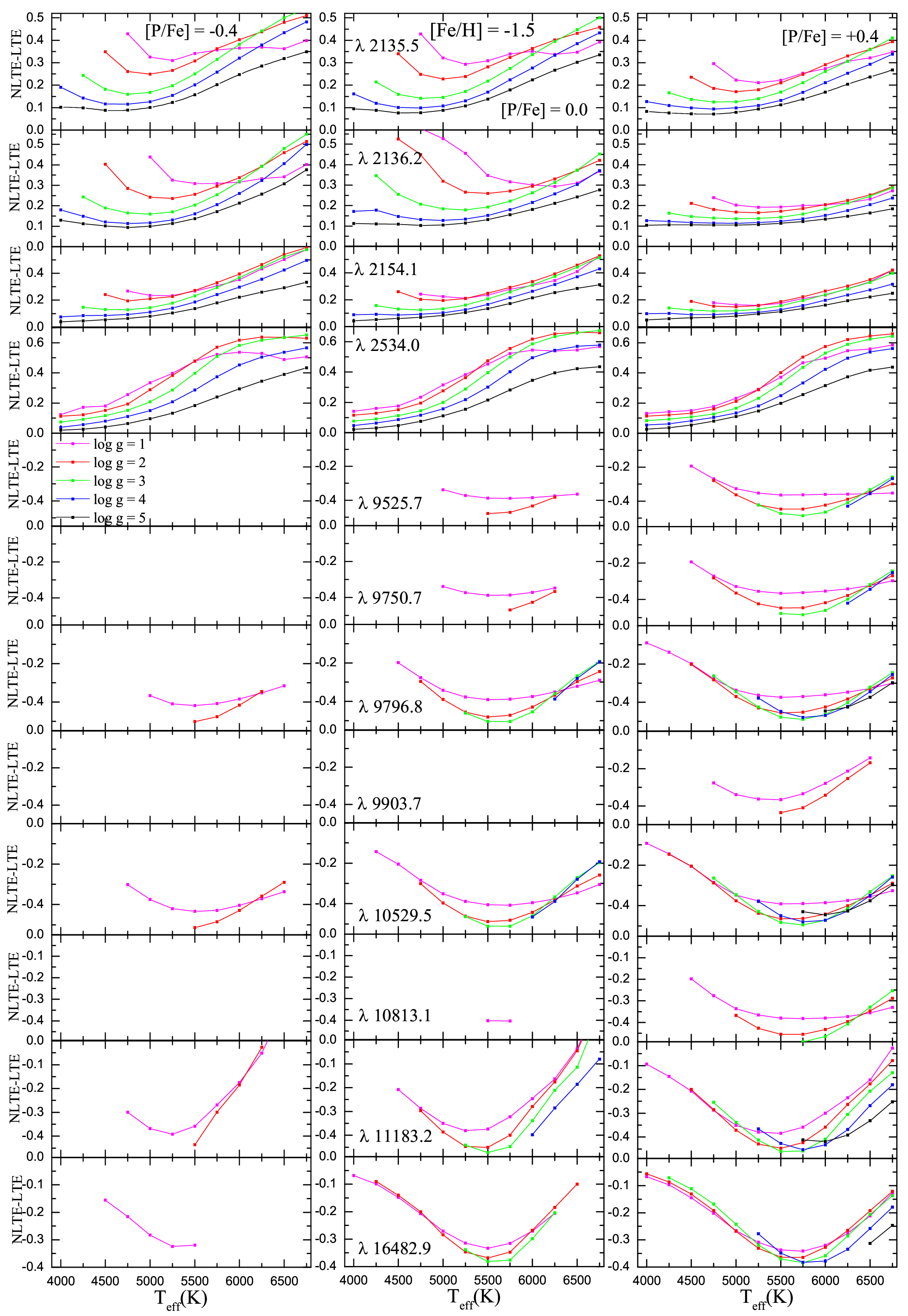}
\caption{Non-LTE corrections for metallicity [Fe/H]=--1.5}
\label{dnlte_m15}
\end{figure*}

\begin{figure*}
\includegraphics[scale=0.5]{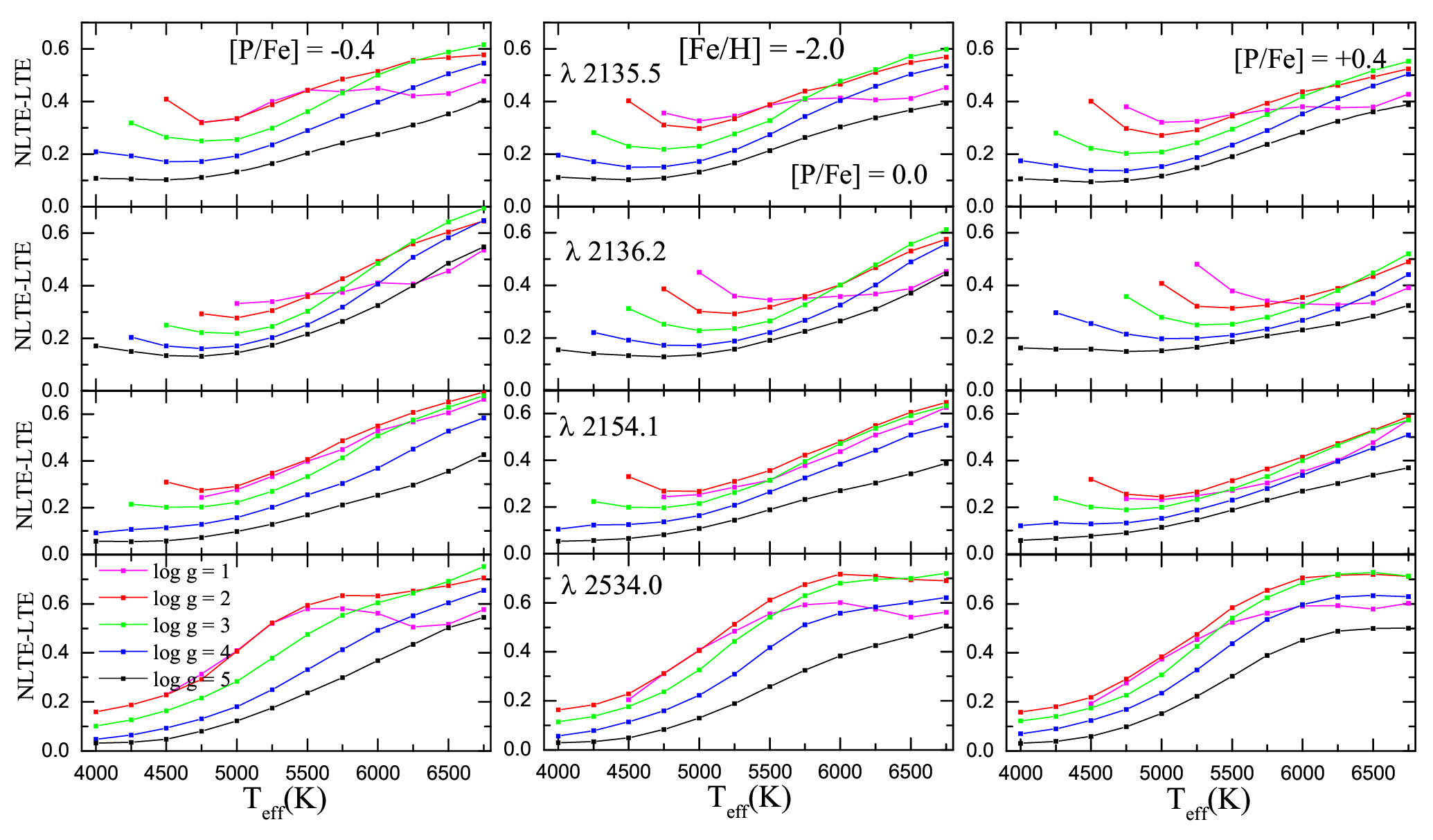}
\caption{Non-LTE corrections for metallicity [Fe/H]=--2.0}
\label{dnlte_m20}
\end{figure*}

\begin{figure*}
\includegraphics[scale=0.5]{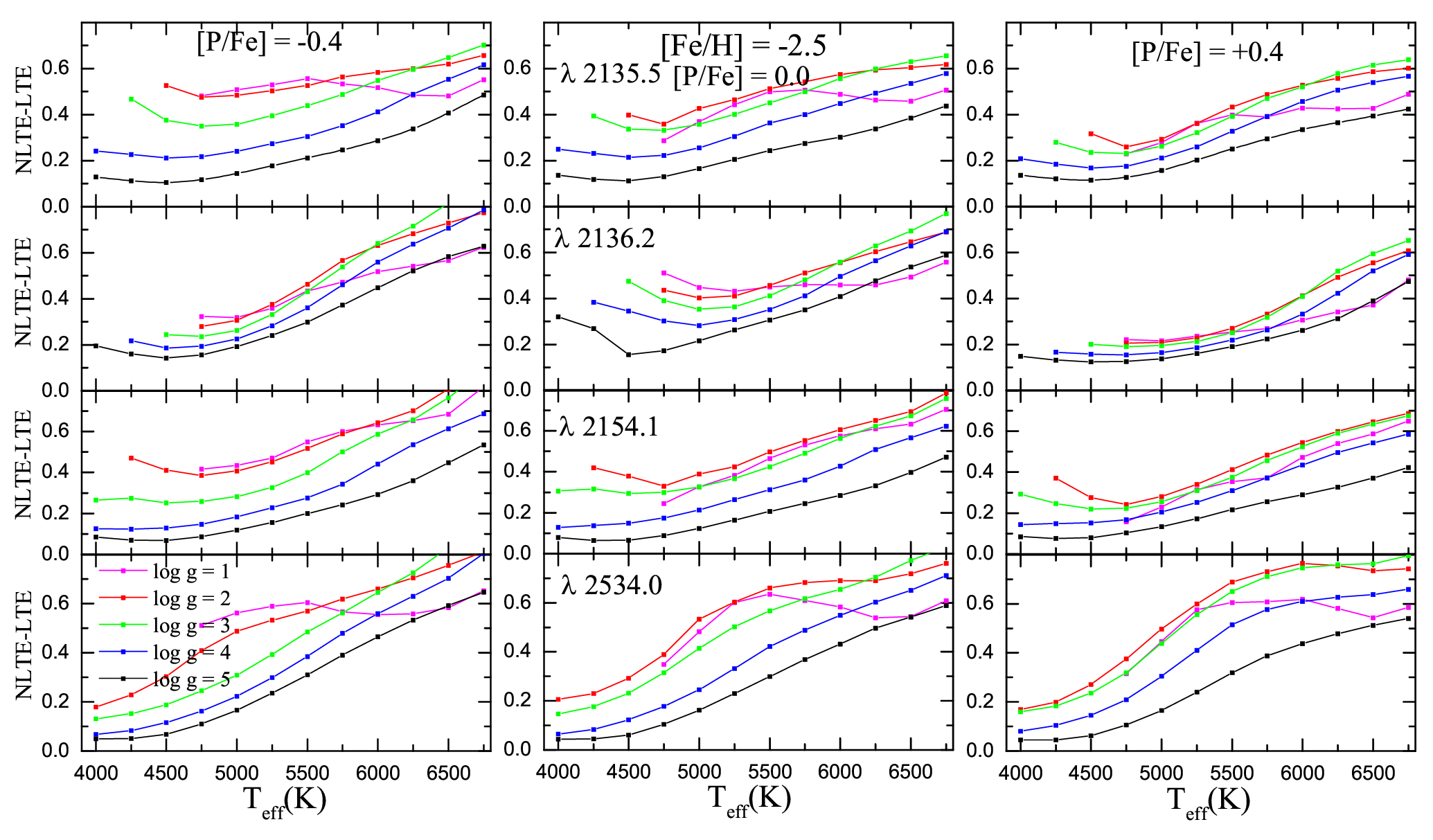}
\caption{Non-LTE corrections for metallicity [Fe/H]=--2.5}
\label{dnlte_m25}
\end{figure*}

\begin{figure*}
\includegraphics[scale=0.5]{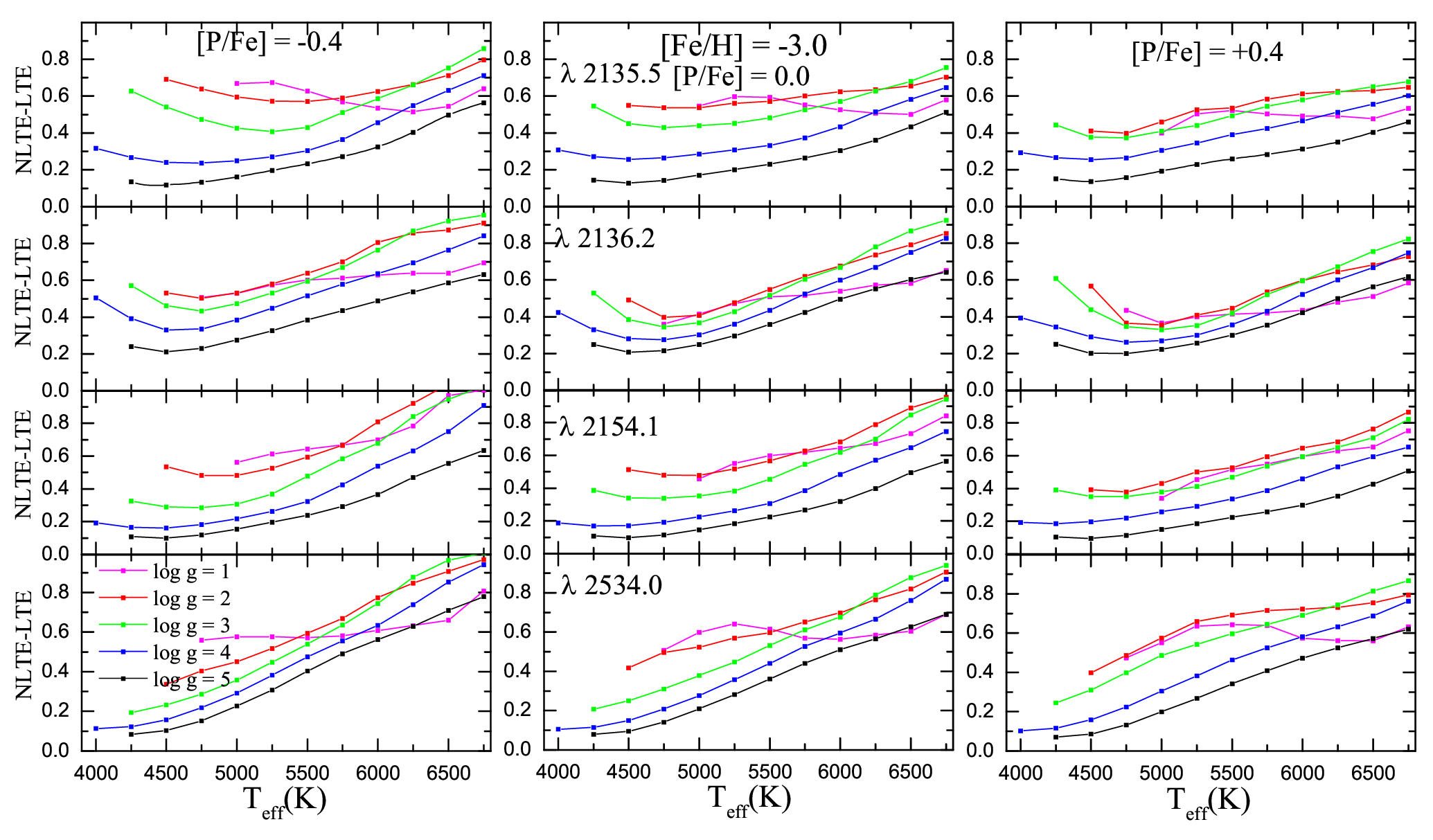}
\caption{Non-LTE corrections for metallicity [Fe/H]=--3.0}
\label{dnlte_m30}
\end{figure*}

\end{appendix}
 
\end{document}